\RequirePackage{fix-cm}
\documentclass[smallextended]{svjour3}       
\smartqed  
\usepackage{graphicx}
\usepackage{amsmath,amssymb,amsfonts}
\usepackage{natbib}
\usepackage{algorithmic}
\usepackage{graphicx}
\usepackage{textcomp}
\usepackage{booktabs}
\usepackage{xcolor}
\usepackage{multirow}
\usepackage{subcaption}
\usepackage{multirow}
\usepackage{listings}
\usepackage{xurl}
\usepackage{hyperref}
\usepackage{balance}
\usepackage{comment}
\usepackage{todonotes}
\usepackage{makecell}
\usepackage{tabularx}
\usepackage{array}
\usepackage{titlesec}
\usepackage{tcolorbox}
\newcommand{\rqbox}[1]{\begin{tcolorbox}[left=4pt,right=4pt,top=2pt,bottom=2pt,colback=gray!5,colframe=gray!40!black,before skip=4pt,after skip=4pt]#1\end{tcolorbox}}

\lstdefinestyle{promptstyle}{
    basicstyle=\ttfamily\scriptsize,
    breaklines=true,
    breakatwhitespace=false,
    frame=single,
    columns=fullflexible,
    backgroundcolor=\color{gray!5},
    xleftmargin=0.5em,
    xrightmargin=0.5em,
    numbers=none,
    showstringspaces=false
}

\newtcbox{\code}{%
on line,
boxsep=1pt,
left=2pt,
right=2pt,
top=1pt,
bottom=1pt,
arc=1.5pt,
boxrule=0pt,
colback=gray!12,
colframe=gray!12,
fontupper=\ttfamily\footnotesize
}

\newtcbox{\sccode}{%
on line,
boxsep=0.5pt,
left=1.5pt,
right=1.5pt,
top=0.5pt,
bottom=0.5pt,
arc=1pt,
boxrule=0pt,
colback=gray!12,
colframe=gray!12,
fontupper=\ttfamily\scriptsize
}

\titleformat{\subsubsection}
  {\normalfont\normalsize\itshape}
  {\alph{subsubsection}.}
  {0.5em}
  {}

\titlespacing*{\subsubsection}
  {0pt}
  {1ex plus .2ex minus .2ex}
  {0.5ex}

\usepackage{xspace}
\newcommand{\Fail}{\textsc{Fail}\xspace}
\newcommand{\Pass}{\textsc{Pass}\xspace}
\newcommand{\Uncompilable}{\textsc{Uncompilable}\xspace}
\newcommand{\UncompilableTimeout}{\textsc{Uncompilable/Timeout}\xspace}
\newcommand{\All}{\textsc{All Cases}\xspace}

\newcommand{\Notpass}{\textsc{Not Pass}\xspace}

\begin{document}

\title{Better Understanding, Better Fixes?
}
\subtitle{A Study of Hallucination in LLM-based Automated Program Repair}

\author{Xuemeng Cai \and
Jiakun Liu \and
Linhan Yang \and
Wei Ma \and
Lingxiao Jiang 
}

\authorrunning{Cai et al.}

\institute{
Xuemeng Cai \and Lingxiao Jiang \at
    School of Computing and Information Systems,
    Singapore Management University, Singapore\\
    \email{xuemengcai@smu.edu.sg}
    \and
Jiakun Liu \and Linhan Yang \at
    Harbin Institute of Technology, Harbin, China
    \and
Wei Ma \at
    Blekinge Institute of Technology, Karlskrona, Sweden
}

\date{Received: date / Accepted: date}

\maketitle

\begin{abstract}

Large language models (LLMs) have significantly advanced automated program repair (APR), yet existing evaluations remain largely result-centric and provide limited insight into hallucination during repair. 
In APR, hallucination may arise not only in final patches but also in the intermediate artifacts that guide patch generation. 
To address this gap, we perform a multi-layered analysis of hallucination throughout the APR process. 
Specifically, we characterize hallucination as the production of patches or intermediate artifacts that are not faithfully grounded in the available repair evidence. 
We examine repair hallucination in final patches and understanding hallucination in intermediate artifacts through three tasks, namely triggering testcase identification, line coverage prediction, and additional testcase generation.
We then evaluate three representative LLMs on 832 Defects4J bugs through automatic evaluation and manual analysis.
Our results show that both repair and understanding hallucinations remain prevalent.
Across models and settings, only 21.0\%--55.9\% of generated patches pass the developer-written test suite. 
Moreover, although more accurate intermediate artifacts are generally associated with successful repairs, this relationship does not always hold. 
Manual analysis of 812 sampled repairs identifies repair hallucinations in 72.7\% of cases, including patches that pass all available tests; incorrect causal localization and incorrect repair strategies account for 45.9\% and 18.5\% of these hallucinations, respectively.
Meanwhile, models frequently misidentify triggering testcases, mispredict line coverage involving branching control flow, and generate additional testcases with missing bug-triggering conditions or incorrect expected behavior. 
Overall, these findings demonstrate that result-centric evaluation alone is insufficient for assessing LLMs' ability to understand and repair bugs, motivating multi-layered evaluation of both final repairs and intermediate artifacts.

\end{abstract}

\section{Introduction}
\label{sec:intro}
Large language models (LLMs) are increasingly being integrated into mainstream software engineering practice. 
Recent studies have demonstrated their applicability to a broad range of tasks, such as code generation, repository-level coding, code summarization, and automated program repair (APR) \citep{chen2021codex,jimenez2024swebench,lomshakov2024proconsul,xia2024chatrepair}. 
Among these tasks, APR has emerged as a prominent application area as it seeks to automatically generate patches for buggy programs, thereby reducing debugging effort and improving software maintenance. 
Compared with earlier search-based, template-based, and learning-based APR techniques \citep{legoues2012genprog,kim2013par,long2015spr,liu2019tbar}, recent LLM-based APR systems have achieved competitive performance on widely used benchmarks by leveraging modern LLMs' capabilities \citep{ribeiro2023llm4apr,xia2024chatrepair,yin2024thinkrepair,bouzenia2025repairagent}.
Despite this progress, APR remains far from a solved problem.
For example, ChatRepair fixes 162 of 337 Defects4J bugs, while ThinkRepair fixes 98 bugs in Defects4J v1.2 \citep{xia2024chatrepair}.
Many generated patches therefore remain incorrect, incomplete, or behaviorally unfaithful.
Moreover, even a patch that passes the available tests may still be semantically incorrect or overfit the test suite \citep{motwani2020quality,petke2024patchoverfitting}.
In this context, hallucination represents an important but underexplored contributor to these failures in LLM-based APR.

Hallucination has become a central concern in LLM-based code intelligence.
Studies of code LLMs show that models may generate syntactically plausible outputs that are semantically unsupported, inconsistent with user requirements, or insufficiently grounded in the surrounding context \citep{Huang2025,zhang2025survey,bang2023multitask,guerreiro2023hallucinations}.
However, most existing studies remain result-centric, analyzing hallucination primarily through the final generated artifact.
In APR, this artifact is the repair patch, namely a candidate fixed version of the buggy code produced by the APR system.
This result-centric focus leaves an important gap because APR requires the model to understand the buggy behavior, reason about the fault, relate the failure to program execution, and finally synthesize a patch \citep{yang2025coast}.
Examining only the final patch therefore cannot reveal whether repair failures stem from misunderstandings of the buggy behavior, execution process, or intended repair semantics.
Consequently, existing result-centric evaluations provide limited insight into whether LLM-based APR systems can faithfully understand and reason about buggy program behavior.

To address this gap, this study moves beyond evaluating APR solely through the final repair patch and performs a multi-layered analysis of hallucination throughout the repair process.
We investigate how hallucination manifests in both final repairs and intermediate artifacts.
In this work, hallucination in APR refers to patches or intermediate artifacts that appear plausible or coherent but are not faithfully grounded in the available repair evidence.
In this work, we define two major forms of hallucination: 
(1) \textit{repair hallucination} refers to hallucinations manifested in the final repair artifact, resulting in patches that fail to compile, deviate from the developer-intended repair semantics, or overfit the available test suite by passing limited tests while violating the intended program behavior; 
and (2) \textit{understanding hallucination} refers to hallucinations manifested in intermediate artifacts, including misidentified bug-triggering testcases, inaccurate line coverage predictions, and invalid additional testcases.
By examining these two forms together, we aim to understand not only what hallucinations appear in APR, but also how hallucinations in the intermediate process relate to the quality of the final repair.

To systematically study understanding hallucinations, we introduce an evaluation framework comprising three complementary APR tasks.
We select these tasks to make the model's otherwise latent program understanding observable through concrete intermediate artifacts and verifiable against execution-grounded evidence.
The tasks form a progression from identifying existing failure evidence, to reasoning about the corresponding program execution, and finally to constructing new behavioral evidence.
\textit{(1) Triggering testcase identification} assesses whether the model can identify the testcases that expose the target bug.
\textit{(2) Line coverage prediction} assesses whether the model can predict the lines executed by the triggering testcases in both the buggy and model-patched programs.
\textit{(3) Additional testcase generation} assesses whether the model can generate new testcases that fail on the buggy program but pass on the developer-written fixed program.
Together, these tasks examine whether patch generation is grounded in faithful program understanding across failure identification, execution reasoning, and behavioral generalization.
Each task also requires the model to generate a repair patch, enabling us to examine the relationship between understanding hallucinations in intermediate artifacts and repair hallucinations in final patches.

Based on this framework, we conduct a large-scale empirical study of three representative LLMs on the Defects4J benchmark, combining automatic evaluation with manual analysis of final patches and intermediate artifacts. 
The goal of this study is not merely to quantify the prevalence of repair and understanding hallucinations, but also to characterize their manifestations across APR tasks and models and to identify their potential contributing factors.

Our results show that both repair and understanding hallucinations are prevalent. 
Among the three models, GPT-5 generally performs best on the understanding tasks, while the line coverage prediction setting achieves the highest plausible-patch rates among the three APR tasks, suggesting that more faithful execution understanding may be associated with better repair outcomes.
At the repair level, only 21.0\%--55.9\% of generated patches pass the developer-written test suite.
Moreover, manual analysis of 812 sampled repairs identifies repair hallucinations in 72.7\% of cases, including patches that pass all available tests; incorrect causal localization and incorrect repair strategies account for 45.9\% and 18.5\% of these hallucinations, respectively.
At the understanding level, the three tasks reveal complementary failure patterns.
In triggering testcase identification, models exactly identify the triggering testcases for only 23.0\%--40.9\% of bugs.
In line coverage prediction, 39 of 44 manually inspected low-scoring cases (88.6\%) involve branch-related control flow, such as \code{if-else} and \code{switch-case} structures.
In additional testcase generation, our manual analysis further reveals that 46 of 185 observed understanding hallucinations (24.9\%) arise because the generated testcases omit necessary bug-triggering conditions.
Across all three tasks, more accurate intermediate artifacts are generally associated with successful repairs, suggesting that faithful program understanding may support patch generation.
However, this relationship is not deterministic, as accurate intermediate artifacts do not always lead to successful repairs, while some passing patches are accompanied by understanding hallucinations.
Together, these results show that the three tasks reveal distinct and complementary forms of understanding hallucination that cannot be captured by patch outcomes alone, demonstrating the value of jointly evaluating final repairs and intermediate artifacts.

These findings also offer practical implications for LLM-based APR.
For example, since intermediate-artifact quality is associated with repair success, task-specific scores could be incorporated into a scoring system for estimating the reliability of generated patches.
Although such scores cannot independently determine repair correctness, they can provide additional signals for identifying potentially unreliable repairs and support more comprehensive patch evaluation.

The remainder of this paper is organized as follows. Section~2 reviews related work; Section~3 presents the research questions and methodology; Section~4 describes the experimental setup; Section~5 reports the results; Section~6 discusses implications and threats to validity; and Section~7 concludes.

\section{Related Work}
\label{sec:related_work}

\subsection{Automated Program Repair}

Automated program repair (APR) aims to automatically generate patches for
buggy programs~\citep{monperrus2018bibliography,legoues2019automated}.
Prior research has explored diverse repair paradigms, including search-based
repair represented by GenProg~\citep{legoues2012genprog}, template-based
repair represented by PAR and TBar~\citep{kim2013par,liu2019tbar},
constraint-based repair represented by SemFix and SPR
~\citep{nguyen2013semfix,long2015spr}, and learning-guided repair represented
by Prophet, CoCoNuT, and SelfAPR~\citep{long2016prophet,lutellier2020coconut,
ye2022selfapr}.
Collectively, these techniques have demonstrated the feasibility of
automatically repairing real-world bugs.

A long-standing challenge in APR is the distinction between plausible and
semantically correct patches.
A patch is commonly considered plausible if it passes the available test suite,
but developer-written tests are sometimes incomplete and cannot fully specify
the intended program behavior~\citep{qi2015plausibility,long2016searchspaces}.
Consequently, patches that pass the test suite may still be semantically
incorrect or overfit the available tests~\citep{smith2015cure,
xin2017identifying,yang2017bettertests,martinez2017automatic,
motwani2020quality,petke2024patchoverfitting}.
Therefore, final test outcomes alone provide incomplete evidence of repair
correctness.

Pretrained code models and large language models (LLMs) have further improved
APR through code infilling, prompt-based generation, retrieval augmentation,
natural-language interaction, and iterative test feedback
~\citep{xia2022alpharepair,joshi2023ring,xia2023plmrepair,
jin2023inferfix,wang2023rapgen,yin2024thinkrepair,xia2024chatrepair,
silva2025repairllama}.
However, existing LLM-based APR studies remain largely patch-centric,
primarily evaluating how many bugs are fixed or whether generated patches pass
the available tests.
Such evaluation provides limited insight into whether a repair is grounded in
a faithful understanding of the underlying buggy behavior.
In contrast, our study examines hallucination in both final repair patches and
task-specific intermediate artifacts, including triggering-testcase
identification, execution-line prediction, and additional testcase generation.

\subsection{Hallucination in LLM-based Code Intelligence}

Hallucination generally refers to model outputs that appear plausible but are
not faithfully grounded in their input or an external source of truth
~\citep{ji2023hallucination,Huang2025}.
In LLM-based code intelligence, hallucinations can manifest as code that is
inconsistent with the available program context, task requirements, APIs, or
program semantics.
Recent studies have investigated this problem specifically in LLM-based code
generation.
For example, Zhang et al. examine hallucinations in practical
repository-level code generation and show that LLMs may produce plausible code
that is insufficiently grounded in the target repository context
~\citep{zhang2025survey}.
Other work constructs taxonomies and benchmarks of code hallucinations,
analyzes defects and API misuse in generated code, and investigates mitigation
through iterative grounding or external documentation
~\citep{eghbali2024dehallucinator,lee2025hallucination,tambon2025bugs}.
These studies establish hallucination as an important threat to the
reliability of LLM-generated code.

However, most existing studies analyze hallucination primarily through the
final generated code, which is insufficient for understanding hallucination
in APR.
Producing a faithful repair typically requires understanding the observed
failure, identifying the tests that expose the bug, localizing its cause,
reasoning about program execution, and synthesizing an appropriate patch
~\citep{wong2016faultlocalization,cheng2025agenticbrt,
cheng2026cogeneration,wu2026tracerepair}.
We similarly expect an LLM-based APR system to ground its repair in these
intermediate understanding and reasoning steps.
Unlike prior work, our study focuses specifically on LLM-based APR and
examines hallucination not only in final patches but also in intermediate
repair artifacts.
This enables us to assess whether apparently coherent repair outputs are
supported by execution-grounded evidence before the final patch is produced.

\subsection{LLM Agents for Automated Program Repair}

Recent work increasingly formulates repository-level issue resolution as an
agentic or structured multi-stage process in which LLMs collect program
context, navigate repositories, invoke tools, execute tests, and iteratively
refine candidate patches~\citep{liu2024agents,yang2024sweagent,
xia2025agentless}.
Representative approaches include autonomous repair systems such as
RepairAgent and AutoCodeRover, as well as systems that improve repair through
cost-aware workflows, repository documentation, or development history
~\citep{bouzenia2025repairagent,autocoderover,patchpilot,
hafixagent,reporepair}.
These systems typically organize issue resolution into stages such as
localization, reproduction, repair, and validation.

Beyond end-to-end repair outcomes, recent studies analyze software engineering
agents at the trajectory level.
For example, prior work examines thought--action--result trajectories to
understand how agents interact with repositories and investigates recurring
reasoning failures and inefficient behaviors that cause agents to go astray
~\citep{bouzenia2025understandingagents,gandhi2025whenagentsgoastray}.
Related work also studies bug reproduction and runtime execution evidence as
explicit components of agentic repair
~\citep{cheng2025agenticbrt,cheng2026cogeneration,wu2026tracerepair}.
These directions demonstrate the value of examining information produced
during repair rather than evaluating only the final submitted patch.

Recent security studies further show that even functionally correct patches
may remain unsafe.
FCV-Attack demonstrates that adversarial inputs can steer code agents toward
functionally correct patches that nevertheless introduce security
vulnerabilities~\citep{peng2025corrects}.
Similarly, SWExploit shows that program repair agents may generate patches
that pass functional tests while still containing exploitable
vulnerabilities~\citep{chen2025redteamingprogramrepair}.
These studies expose important security limitations of patch-level
correctness, but focus primarily on adversarially induced vulnerabilities in
final patches.

Our study examines a complementary dimension of APR reliability by evaluating
whether final patches and task-specific intermediate artifacts are
semantically faithful to execution-grounded evidence.
This matters because a plausible patch or coherent repair process may still
rest on an incorrect understanding of the underlying bug behavior.
Unlike trajectory-level studies that focus on the coherence and efficiency of
complete agent trajectories, or security studies that examine adversarially
induced vulnerabilities, we study hallucination in general, non-adversarial
APR settings.

\section{Methodology}
\label{sec:methodology}
This section presents the methodology of our study and is organized as follows.
First, we introduce the conceptual organization and overall workflow of the study.
Second, we formulate the research questions.
Third, we describe the design and automatic evaluation of three complementary APR tasks.
Finally, we present the human annotation procedure used to characterize hallucination types and their potential contributing factors.

\subsection{Overview}
\label{sec:study-overview}

\begin{figure*}[t]
    \centering
    \includegraphics[width=0.95\textwidth]{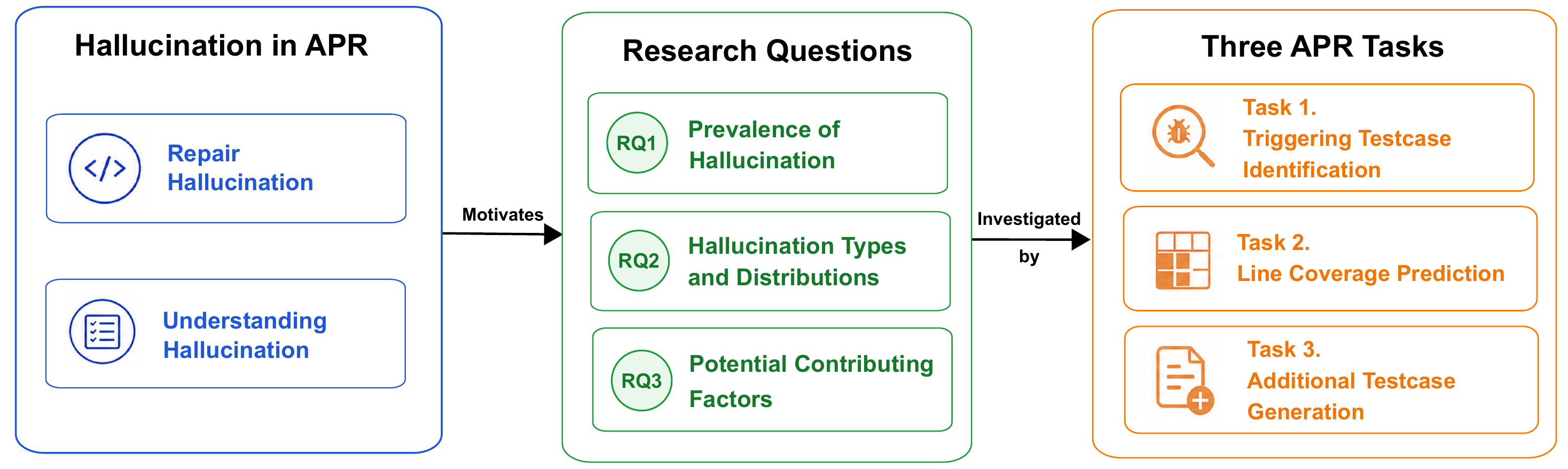}
    \caption{Conceptual overview of our study on hallucination in LLM-based APR.}
    \label{fig:hallucination-overview}
\end{figure*}

Figure~\ref{fig:hallucination-overview} illustrates the conceptual organization of our study.
We distinguish between two forms of hallucination in APR.
\textit{(1) Repair hallucination} occurs in the final generated patch and refers to a patch that does not implement a valid or developer-intended repair.
This includes patches that fail to compile, fail the developer-written test suite, or pass the available tests but overfit them or deviate from the intended repair semantics.
\textit{(2) Understanding hallucination} occurs in an intermediate artifact and refers to model-generated information that is inconsistent with execution-grounded evidence about the buggy or repaired program.
Such artifacts indicate that the model misunderstands the failure evidence, program execution behavior, or intended repair semantics.

Based on these two forms of hallucination, we formulate three research questions concerning hallucination prevalence, hallucination types and distributions, and potential contributing factors.
To answer these research questions while making otherwise latent program understanding observable and verifiable, we design three complementary APR tasks.
These tasks form a progression from identifying existing failure evidence, to reasoning about the corresponding program execution, and finally to constructing new behavioral evidence.

\textit{(1) Triggering testcase identification} examines whether an LLM can identify the testcases that expose the target bug.
In APR, triggering testcases provide the primary failure evidence for understanding buggy behavior and guiding patch generation.
If the model identifies unrelated testcases as bug-triggering or fails to identify actual triggering testcases, its subsequent repair may be based on an incorrect understanding of the failure.
Accordingly, we define an understanding hallucination in this task as any prediction that omits at least one actual triggering testcase or incorrectly identifies a non-triggering testcase as a trigger.

\textit{(2) Line coverage prediction} examines whether an LLM can identify which lines are executed when the triggering testcases run on both the buggy and model-patched programs.
In APR, execution traces provide important evidence for understanding how a failure is triggered and how a repair changes program behavior.
If the model omits lines that are actually executed or predicts unexecuted lines as executed, it may misunderstand the dynamic behavior of the buggy or model-patched program.
Accordingly, we define an understanding hallucination in this task as any prediction that does not exactly match the observed executed-line set for either program version.

\textit{(3) Additional testcase generation} examines whether an LLM can generate new testcases that expose the target bug and are consistent with the developer-intended fixed behavior.
In APR, additional testcases can provide extra behavioral evidence beyond the given triggering testcases.
However, generated testcases may also reinforce an incorrect repair: a testcase may agree with the model-generated patch while still being inconsistent with the developer-written patch.
Such cases are especially problematic because the generated patch and testcase appear mutually consistent, but together encode repair semantics that deviate from the developer-intended behavior.
Therefore, this task evaluates generated testcases against both the developer-written fixed program and the model-generated patched program.
We define an understanding hallucination in this task as a generated testcase that fails to expose the bug in the original program or is inconsistent with the behavior of the developer-written fixed program.

Together, the three tasks examine whether the model can identify failure evidence, reason about its dynamic execution, and generalize the inferred behavior into new testcases.
Each task also requires the model to generate a repair patch, enabling paired analysis of understanding hallucinations in intermediate artifacts and repair hallucinations in final patches.

\begin{figure*}[t]
    \centering
    \includegraphics[width=0.9\textwidth]{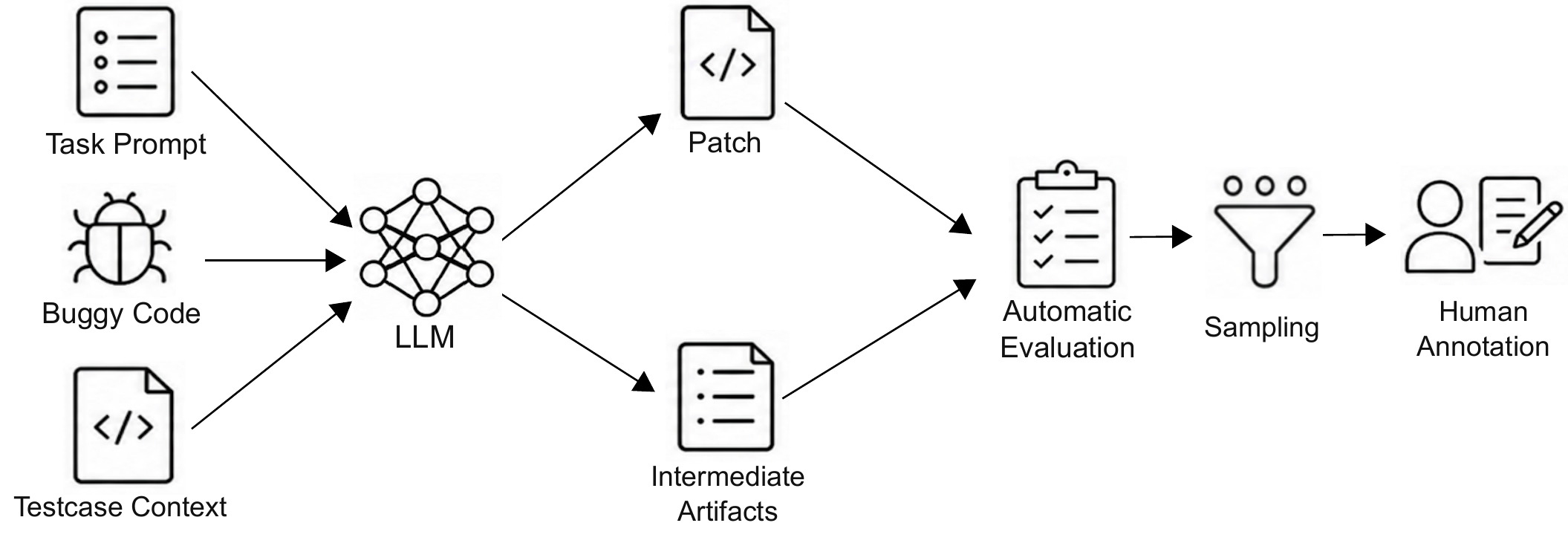}
    \caption{Overall workflow of our study.}
    \label{fig:methodology-workflow}
\end{figure*}

Figure~\ref{fig:methodology-workflow} presents the overall workflow of our study.
For each APR task, the LLM receives a task-specific prompt, buggy code, and testcase context, and produces a repair patch together with one or more task-specific intermediate artifacts.
We automatically evaluate the generated patch and intermediate artifacts using developer-written tests.
We then sample the outputs for human annotation.
The automatic evaluation provides evidence for quantifying the prevalence of repair and understanding hallucinations, while the human annotation supports the characterization of their types, distributions, and potential contributing factors.
The following sections describe the task-based evaluation framework and human annotation procedure in detail.

\subsection{Research Questions}

We formulate three research questions from three complementary perspectives: hallucination prevalence, hallucination types and distributions, and potential contributing factors.

\textbf{RQ1: How often do LLMs exhibit hallucinations when performing different APR tasks?}
Through this RQ, we quantify the prevalence of repair hallucinations in final patches and understanding hallucinations in task-specific intermediate artifacts across APR tasks and models.

\textbf{RQ2: What types of repair and understanding hallucinations occur when LLMs perform APR tasks, and how are these types distributed?}
Through this RQ, we characterize the concrete manifestations of repair hallucinations and understanding hallucinations and analyze their distributions across APR tasks and models.

\textbf{RQ3: What potential factors may contribute to hallucinations when LLMs perform APR tasks?}
Through this RQ, we identify potential contributing factors based on the hallucination patterns and empirical evidence observed in our analysis.



\subsection{Triggering Testcase Identification}
\label{sec:trigger-testcase-identification}

\begin{figure*}[t]
    \centering
    \includegraphics[width=0.8\textwidth]{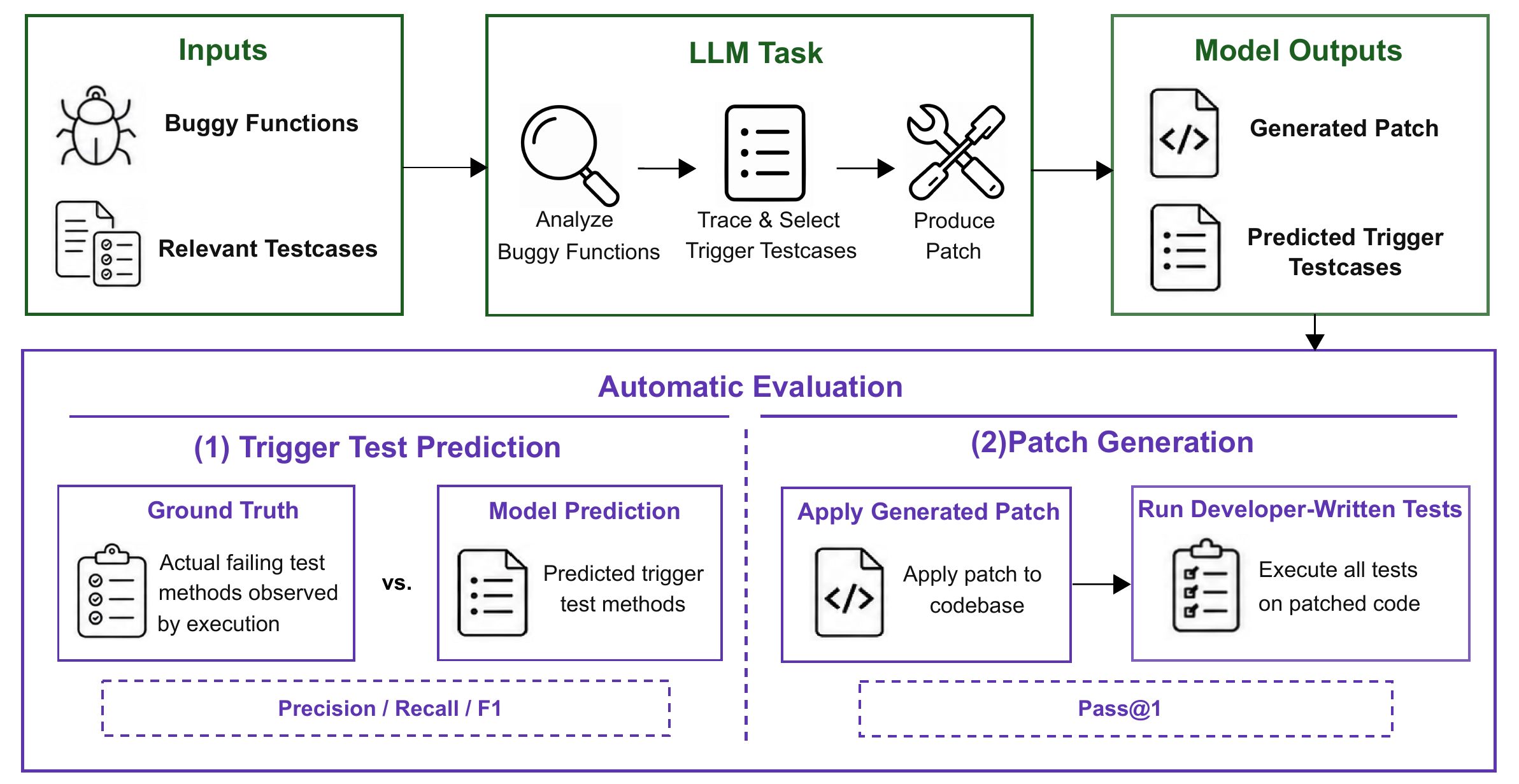}
    \caption{Evaluation process for triggering testcase identification.}
    \label{fig:task1-overview}
\end{figure*}


\subsubsection{Task Design}

Figure~\ref{fig:task1-overview} presents our evaluation process for the triggering testcase identification task. 
Given a bug, we provide the model to be evaluated with the buggy functions and all relevant developer-written testcases. 
Here, a \textit{relevant testcase} refers to any test that reaches at least one class modified by the ground-truth patch during execution.
Formally, for each bug \(b\), let \(F_b\) denote the set of buggy functions provided to the model, and let \(R_b = \{r_1, r_2, \dots, r_m\}\) denote the set of relevant test methods. 
A test method refers to an individual developer-written JUnit test recognized by the testing framework, such as a method annotated with \code{@Test}.
The model is asked to identify which test methods actually trigger the bug and to generate a repair patch. 
Accordingly, the model produces two outputs: a predicted set of triggers \(\hat{T}_b \subseteq R_b\) and a generated repair patch \(\hat{P}_b\).

\subsubsection{Automatic Evaluation}
\label{sec:automatic_evaluation}

The automatic evaluation of this task covers both the intermediate artifact and the final repair output.
To evaluate the set of predicted triggering testcases, we first obtain the ground-truth triggering testcases by executing all relevant developer-written testcases on the original buggy version.
The test methods that fail on the buggy version are treated as the ground-truth triggering testcases, denoted as \(T_b\).
We then compare the model-predicted triggering testcases \(\hat{T}_b\) with \(T_b\).
Specifically, we compute precision, recall, and F1 score:
\[
\mathrm{Precision} = \frac{|\hat{T}_b \cap T_b|}{|\hat{T}_b|},
\]
\[
\mathrm{Recall} = \frac{|\hat{T}_b \cap T_b|}{|T_b|},
\]
\[
\mathrm{F1} = \frac{2 \times \mathrm{Precision} \times \mathrm{Recall}}{\mathrm{Precision} + \mathrm{Recall}}.
\]


To evaluate the generated repair patch, we apply the patch \(\hat{P}_b\) to the original codebase and run the developer-written test suite on the patched version. 
We denote the patch evaluation result as \(T_{\text{patch}} \in \{\Pass, \Notpass, \\ 
\Uncompilable\}\). 
A patch is labeled as \Pass \space if all developer-written tests pass, \Notpass \space if it compiles but fails at least one test, and \Uncompilable \space if the patched version fails to compile. 
We treat \Notpass \space and \Uncompilable \space patches as repair hallucinations because they fail to produce a valid repair under the developer-written test suite.
However, passing the test suite does not necessarily establish semantic correctness, as a \Pass \space patch may overfit the available tests or deviate from the developer-intended repair semantics.
Such hallucinations cannot be reliably identified from test outcomes alone.
We therefore complement the automatic evaluation with human annotation, in which sampled \Pass \space patches are compared with the developer-written patches and available execution evidence to determine whether they implement the intended repair semantics.
The detailed annotation procedure is described in Section~\ref{sec:human-annotation}.

\subsection{Line Coverage Prediction}
\label{sec:execution-line-prediction}

\begin{figure*}[t]
    \centering
    \includegraphics[width=0.8\textwidth]{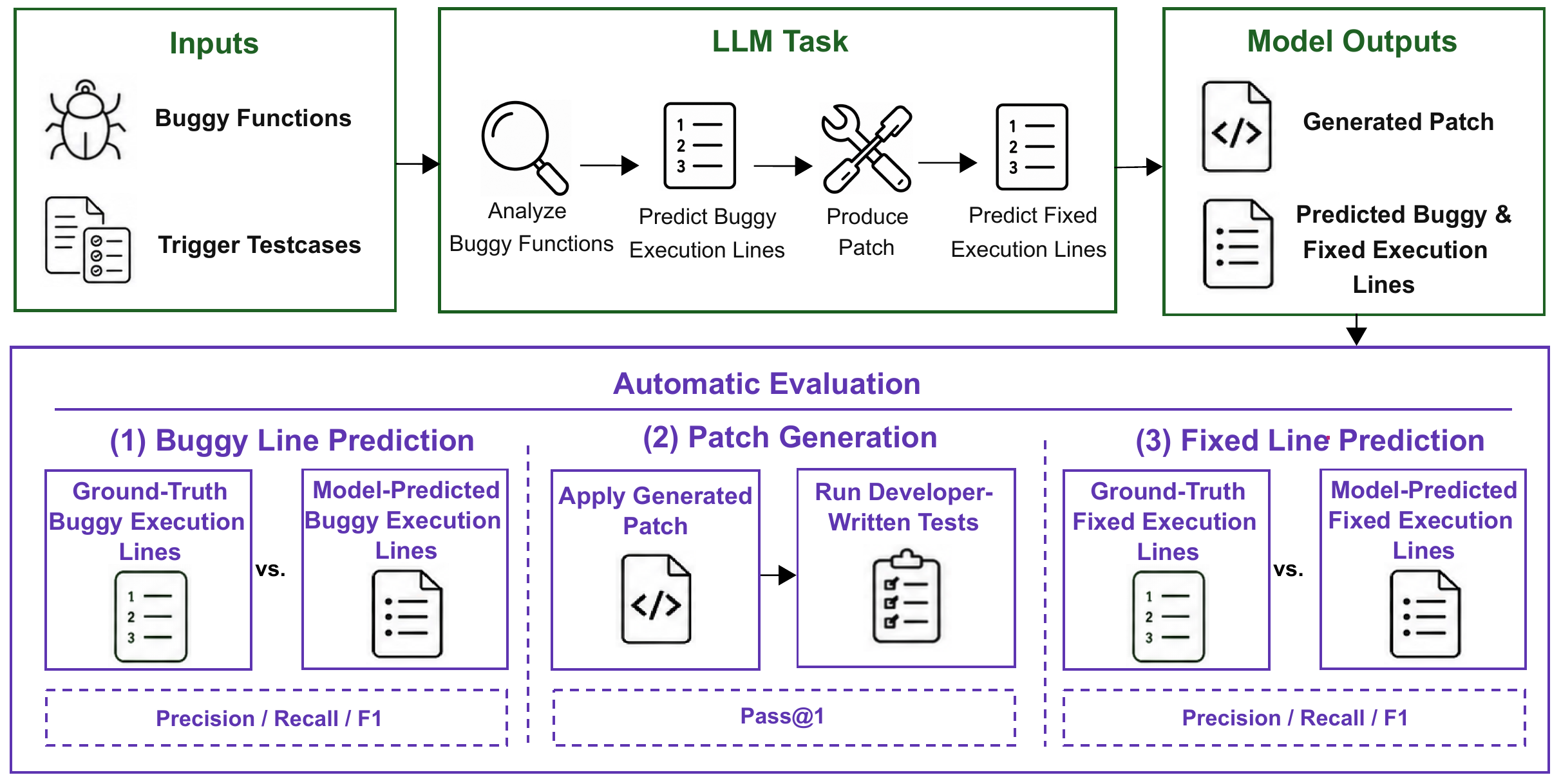}
    \caption{Evaluation process for line coverage prediction.}
    \label{fig:task2-overview}
\end{figure*}


\subsubsection{Task Design}

Figure~\ref{fig:task2-overview} presents the evaluation process of the line coverage prediction task.
Given a bug, we provide the model with the buggy functions and all triggering testcases.
The model is first asked to analyze the execution behavior of the buggy functions under the triggering testcases and predict which lines in these functions are executed.
It then generates a repair patch and predicts which lines in the corresponding model-patched functions would be executed under the same triggering testcases.

Formally, for each bug \(b\), let \(F_b = \{f_1, f_2, \dots, f_n\}\) denote the set of buggy functions provided to the model, and let \(T_b\) denote the set of all triggering testcases.
The model produces three outputs: a generated repair patch \(\hat{P}_b\), a set of lines predicted to be executed in the buggy functions \(\hat{L}^{\text{bug}}_b\), and a set of lines predicted to be executed in the model-patched functions after applying \(\hat{P}_b\), denoted as \(\hat{L}^{\text{fix}}_b\).
Here, the model-patched functions refer to the functions obtained from the model-generated patch.

\subsubsection{Automatic Evaluation}

To evaluate the model's prediction for the buggy program, we execute all triggering testcases \(T_b\) on the original buggy program and use an instrumentation tool to collect the lines executed in the buggy functions.
We exclude lines that are not informative for line coverage evaluation, including blank lines, comment-only lines, and lines containing only braces.
Let \(\mathcal{L}^{\text{bug}}_b = \{\ell_1, \ell_2, \dots, \ell_m\}\) denote the set of remaining lines in the buggy functions after filtering.
The lines executed in the buggy functions under the triggering testcase set \(T_b\) are denoted as \(L^{\text{bug}}_b(T_b) \subseteq \mathcal{L}^{\text{bug}}_b\).
We apply the same filtering rule to the model prediction and denote the remaining lines predicted to be executed under \(T_b\) as \(\hat{L}^{\text{bug}}_b(T_b) \subseteq \mathcal{L}^{\text{bug}}_b\).

We compare the predicted and observed executed-line sets using precision, recall, and F1 score:
\[
\mathrm{Precision}^{\text{bug}}_b =
\frac{
\left|L^{\text{bug}}_b(T_b) \cap \hat{L}^{\text{bug}}_b(T_b)\right|
}{
\left|\hat{L}^{\text{bug}}_b(T_b)\right|
},
\]
\[
\mathrm{Recall}^{\text{bug}}_b =
\frac{
\left|L^{\text{bug}}_b(T_b) \cap \hat{L}^{\text{bug}}_b(T_b)\right|
}{
\left|L^{\text{bug}}_b(T_b)\right|
},
\]
\[
\mathrm{F1}^{\text{bug}}_b =
\frac{
2 \times \mathrm{Precision}^{\text{bug}}_b \times \mathrm{Recall}^{\text{bug}}_b
}{
\mathrm{Precision}^{\text{bug}}_b + \mathrm{Recall}^{\text{bug}}_b
}.
\]
Precision measures the proportion of predicted execution lines that are actually executed, while recall measures the proportion of actually executed lines identified by the model.
We use F1 as the primary metric because it balances incorrectly predicted and omitted execution lines without rewarding correctly predicted unexecuted lines.
If \(L^{\text{bug}}_b(T_b) \neq \emptyset\) and \(\hat{L}^{\text{bug}}_b(T_b) = \emptyset\), we set precision, recall, and F1 to 0.
If both sets are empty, we treat the prediction as fully correct and set all three metrics to 1.

For the generated patch \(\hat{P}_b\), we use the same patch-evaluation procedure described in Section~\ref{sec:trigger-testcase-identification}, which classifies the patch as \Pass, \Notpass, or \Uncompilable \space based on the developer-written test suite.

Finally, we evaluate the model's prediction for the model-patched program.
For each compilable generated patch, we execute the same triggering testcase set \(T_b\) on the program obtained by applying \(\hat{P}_b\) and collect the lines executed in the model-patched functions.
We apply the same filtering rule and let \(\mathcal{L}^{\text{fix}}_b\) denote the remaining lines in the model-patched functions.
The observed and predicted sets of executed lines are denoted as \(L^{\text{fix}}_b(T_b) \subseteq \mathcal{L}^{\text{fix}}_b\) and \(\hat{L}^{\text{fix}}_b(T_b) \subseteq \mathcal{L}^{\text{fix}}_b\), respectively.
We compute \(\mathrm{Precision}^{\text{fix}}_b\), \(\mathrm{Recall}^{\text{fix}}_b\), and \(\mathrm{F1}^{\text{fix}}_b\) using the same definitions.
If the generated patch is \Uncompilable, the model-patched program cannot be executed, and we set \(\mathrm{Precision}^{\text{fix}}_b\), \(\mathrm{Recall}^{\text{fix}}_b\), and \(\mathrm{F1}^{\text{fix}}_b\) to 0.

\subsection{Additional Testcase Generation}
\label{sec:additional-test-generation}

\begin{figure*}[t]
    \centering
    \includegraphics[width=0.8\textwidth]{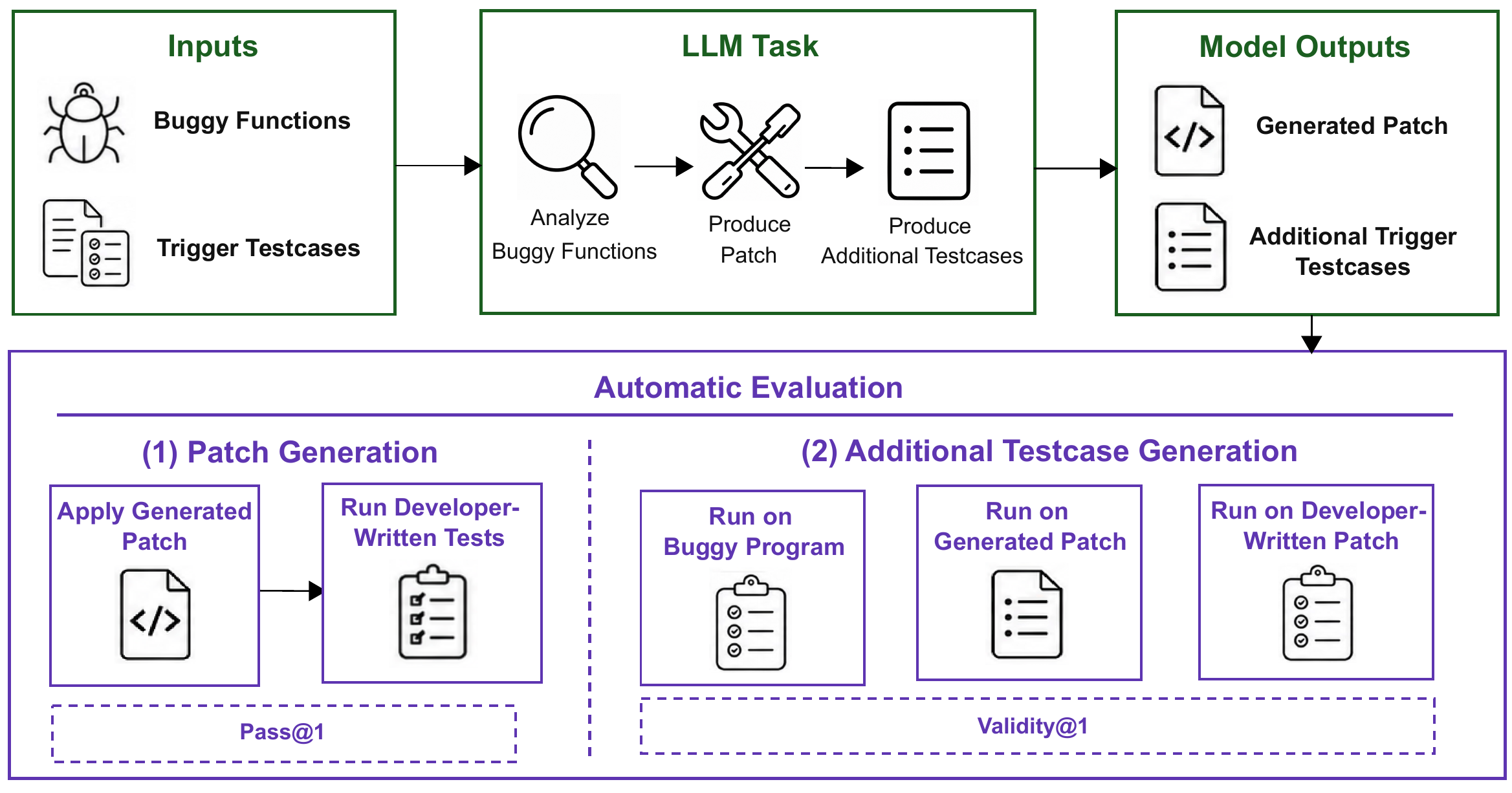}
    \caption{Evaluation process for additional testcase generation.}
    \label{fig:task3-overview}
\end{figure*}

\subsubsection{Task Design}
Figure~\ref{fig:task3-overview} presents the overall workflow of the additional testcase generation task.
Given a bug, we provide the model with the buggy functions and all triggering testcases.
The model is asked to first generate a repair patch and then generate additional triggering testcases.
Each generated testcase is expected to fail on the original buggy version and pass on the fixed version.

Formally, for each bug \(b\), let \(F_b = \{f_1, f_2, \dots, f_n\}\) denote the set of buggy functions provided to the model, and let \(T_b\) denote the set of triggering testcases.
The model produces two outputs: a generated repair patch \(\hat{P}_b\) and a set of generated additional testcases \(\hat{A}_b = \{\hat{a}_1, \hat{a}_2, \dots, \hat{a}_k\}\).
Each generated testcase \(\hat{a}_i\) includes a test method in the form of a Java method recognized by the testing framework, such as a method annotated with \code{@Test}.
It also specifies the project-relative path of the test file into which the generated method should be inserted.
This path must exactly match the test file path of one of the provided triggering testcases, so that the generated method can be inserted into an existing test file and executed within the existing testing environment.

\subsubsection{Automatic Evaluation}

The automatic evaluation of this task examines both the generated repair patch and the generated additional testcases.

For the generated patch \(\hat{P}_b\), we use the same evaluation procedure as described in Section~\ref{sec:trigger-testcase-identification}. We use \(T_{\text{patch}}\) to denote the outcome of the model-generated patch on the developer-written test suite.

To evaluate the generated additional triggering testcases, we execute each generated testcase on three program versions: the original buggy program, the developer-written fixed program, and the model-generated patched program.
The original buggy program is used to check whether the generated testcase can expose the target bug.
Unlike the preceding tasks, this task requires an external oracle for the expected fixed behavior of model-generated triggering testcases.
We therefore execute each testcase on the developer-written fixed program, which serves as the behavioral oracle for the intended fix.
This check prevents a testcase from being treated as valid merely because it agrees with the model's own repair, which may itself encode incorrect repair semantics.
Finally, the model-generated patched program is used to examine whether the generated testcase is consistent with the model's own repair.

For each generated testcase \(\hat{a}_i \in \hat{A}_b\), let \(T_{\text{buggy}}(\hat{a}_i)\),\(T_{\text{gt}}(\hat{a}_i)\), and  \(T_{\text{model}}(\hat{a}_i)\) denote its execution outcome on the original buggy program, the developer-written fixed program, and the model-generated patched program, respectively.

A generated testcase \(\hat{a}_i \in \hat{A}_b\) is considered valid if it fails on the original buggy program and passes on the developer-written fixed program, i.e.,
\[
T_{\text{buggy}}(\hat{a}_i) = \Fail
\quad \wedge \quad
T_{\text{gt}}(\hat{a}_i) = \Pass.
\]
We treat a generated testcase as invalid in any of the following cases:
\begin{itemize}
\item it does not provide a valid Java test method and its corresponding test-file path;
\item it cannot be compiled or executed on either the original buggy program or the developer-written fixed program;
\item it passes on the original buggy program, indicating that it does not expose the target bug;
\item it fails on the developer-written fixed program, indicating that it is inconsistent with the developer-intended fixed behavior.
\end{itemize}




\subsection{Human Annotation and Evaluation}
\label{sec:human-annotation}

\begin{figure*}[t]
    \centering
    \includegraphics[width=0.85\textwidth]{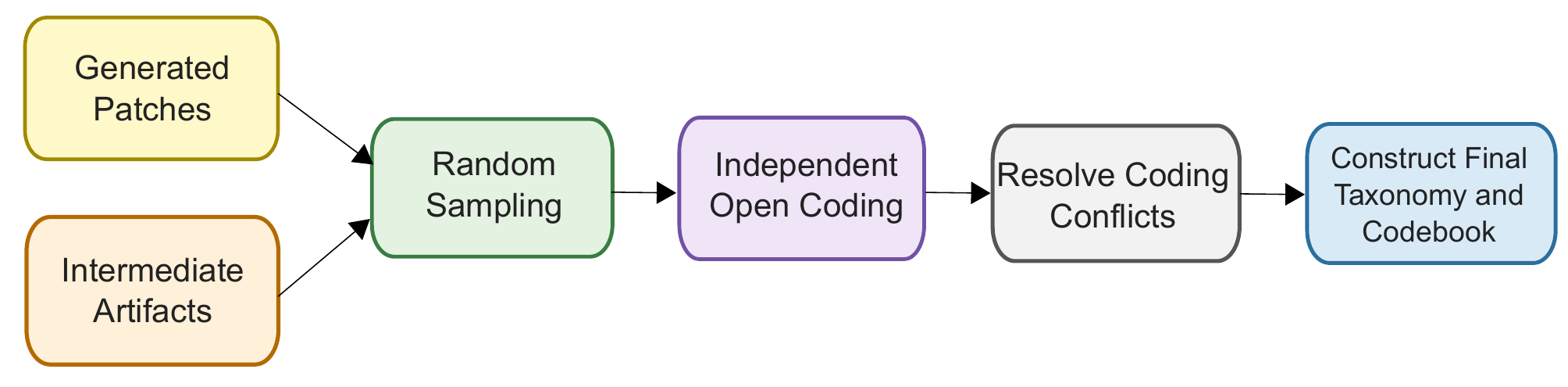}
    \caption{Human annotation workflow of our study.}
    \label{fig:human-annotation}
\end{figure*}

After automatic evaluation, we conduct human annotation to characterize the forms and potential causes of hallucinations.
Automatic evaluation determines whether a generated artifact satisfies execution-based criteria, but it does not explain how the hallucination manifests or why the output is incorrect.
We therefore manually inspect sampled generated patches and task-specific intermediate artifacts.
Figure~\ref{fig:human-annotation} presents the workflow of our annotation procedure.

\subsubsection{Data Sampling}

We sample cases from both generated patches and task-specific intermediate artifacts.
Since our study evaluates three APR tasks and three LLMs, we stratify the evaluated outputs by task and model, resulting in nine task-model groups.
For each group, we sample at least the minimum number of cases required for a 95\% confidence level and a 10\% margin of error.

Each sampled case includes both the generated patch and the corresponding intermediate artifact, allowing annotators to examine repair hallucination and understanding hallucination for the same model output.
The detailed population and sample sizes are reported in Section~\ref{sec:experiment:implementation}.

\subsubsection{Independent Open Coding}

We use independent open coding to identify hallucination patterns.
Two authors with 3--5 years of Java programming experience independently inspect each sampled case and assign descriptive codes.
Rather than using a predefined taxonomy, annotators record the concrete failure manifestation, execution-based evidence, and likely cause of the incorrect output.

For generated patches, annotators first examine compilation errors for \(\Uncompilable\) patches.
For compilable patches, they inspect the developer-written patch to understand the intended repair location and semantics, and then compare it with the model-generated patch to identify its modified locations and inferred repair intention.
For \(\Notpass\) patches, annotators further inspect the failing developer-written testcases to understand how the model patch deviates from the expected behavior.
For \(\Pass\) patches, they compare the model-generated patch with the developer-written patch to determine whether it implements the intended repair semantics or merely overfits the available tests.

For task-specific intermediate artifacts, annotators compare the model output with the corresponding execution-grounded oracle.
For triggering testcase identification, the discrepancy between the predicted triggering testcase set \(\hat{T}_b\) and the ground-truth triggering testcase set \(T_b\) can be automatically characterized by set relations, so we do not manually annotate this intermediate artifact.
For line coverage prediction, annotators inspect the mismatched regions between predicted execution lines and instrumentation-collected execution lines, and label the main code structures associated with the prediction error.
For additional testcase generation, annotators first check whether the generated testcase can be compiled and executed.
If it is \(\Uncompilable\), they inspect the compilation errors.
For executable but invalid testcases, they analyze the testcase content, execution path, assertion oracle, and execution outcomes on the buggy, developer-written fixed, and model-generated patched programs.

\subsubsection{Codebook Construction}

After independent open coding, we consolidate the descriptive codes into a preliminary codebook.
We merge similar codes, define each category, specify labeling criteria and required evidence, and record representative corner cases to clarify category boundaries.
The preliminary codebook is then used for label comparison and conflict resolution.

\subsubsection{Conflict Resolution}

Using the preliminary codebook, we compare the labels assigned by the two annotators and measure the initial inter-annotator agreement.
We use Cohen's kappa \((\kappa)\) because it accounts for agreement occurring by chance.
We compute agreement separately for repair hallucination labels and understanding hallucination labels.
The initial agreement is \(\kappa=0.82\) for repair hallucination annotations and \(\kappa=0.49\) for understanding hallucination annotations.

We resolve coding conflicts through face-to-face review sessions involving the two annotators and two additional authors.
The two additional authors have over nine years and over fifteen years of software engineering research experience, respectively.
The final label is assigned only after the annotators reach agreement.

\subsubsection{Taxonomy Construction}

After conflict resolution, we finalize the hallucination taxonomy based on the resolved labels and updated codebook.
We then apply the finalized taxonomy to all sampled cases and summarize the category distributions across tasks and models.
These taxonomy results are used to characterize hallucination forms and support our analysis of potential contributing factors.

\section{Experimental Setup}
\label{sec:experiment}
This section presents the experimental setup of our study. 
We first introduce the models evaluated in our experiments, then describe the benchmark dataset and its statistics. 
Next, we explain the implementation of the three APR tasks, including prompt construction, output parsing, patch application, execution-line collection, and generated testcase execution. 
We also report the number of generated outputs and sampled cases used for human annotation. 
Finally, we describe the experimental configuration, including model API settings, execution tools, timeout settings, and the Java environment.

\subsection{Studied Models}
\label{sec:experiment:models}

To support our task design, models are required to have strong capabilities in code generation, program reasoning, and instruction following.
Accordingly, we evaluate three representative LLMs: GPT-5, DeepSeek-R1, and Claude Sonnet 4.5.
At the beginning of our experiments in November 2025, these models were competitive and widely accessible through their corresponding provider APIs.
As discussed in Section~\ref{sec:threats-validity}, although newer models may change the absolute performance numbers, our methodology and qualitative findings remain applicable to future LLM-based APR systems.

\textbf{GPT-5}\citep{gpt5} is a proprietary general-purpose foundation model developed by OpenAI.
It demonstrates strong capabilities in reasoning, code generation, and instruction following, making it suitable for our APR tasks.

\textbf{DeepSeek-R1}\citep{deepseekr1} is an open-weight reasoning model developed by DeepSeek.
It is optimized for reasoning-intensive tasks and has shown strong performance in code-related benchmarks.

\textbf{Claude Sonnet 4.5}\citep{claude45} is a proprietary model developed by Anthropic.
It is designed for strong performance in coding, agentic tasks, and long-context reasoning, which aligns with the requirements of our study.




\subsection{Dataset}
\label{sec:experiment:dataset}

\begin{figure*}[t]
    \centering
    \includegraphics[width=0.9\textwidth]{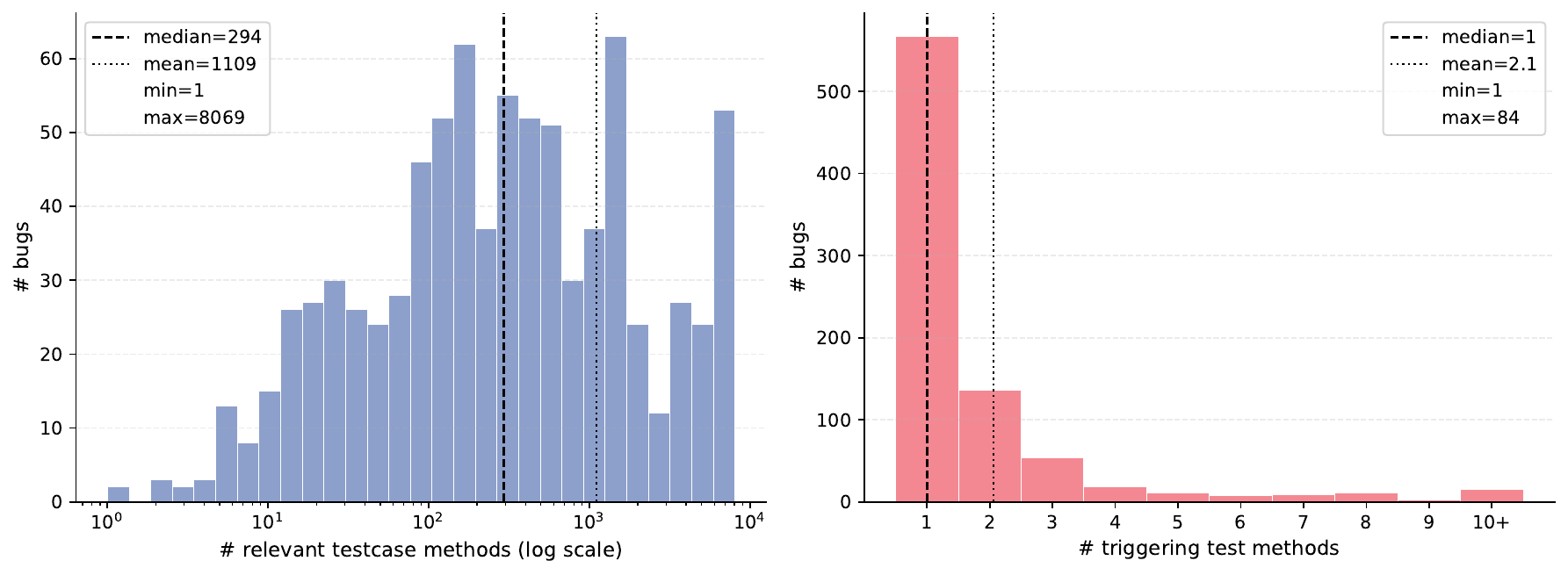}
    \caption{Distribution of relevant test methods and triggering test methods per bug.}
    \label{fig:testcase-method-dist}
\end{figure*}

We conduct our experiments on \textsc{Defects4J}\citep{defects4j}, a widely used benchmark for Java automated program repair.
Each bug in \textsc{Defects4J} contains a buggy program version, a developer-written fixed version, and a developer-written test suite.
These artifacts provide the necessary ground truth for evaluating both generated repair patches and task-specific intermediate artifacts in our study.

We then filter the bugs according to the requirements of our task design.
(1) The buggy version and the developer-written fixed version must be successfully checked out, compiled, and tested in our environment;
(2) each bug must have at least one triggering testcase, i.e., at least one developer-written test method fails on the buggy version;
and (3) for each bug, every developer-modified location must be contained within an identifiable function, and the patch must not consist solely of inserting an entire new function, because our task inputs include only extracted buggy functions rather than full source files.
After filtering, our dataset contains 832 bugs.

For each selected bug, we extract all developer-modified functions from the buggy version and use them as the buggy program context provided to the model.
We collect relevant developer-written test methods as the testcase context, where a testcase is considered relevant if its execution loads at least one class containing a developer-modified function.
Among these relevant testcases, the test methods that fail on the buggy version are treated as triggering testcases.
Thus, for each bug \(b\), we construct the buggy function set \(F_b\), the relevant testcase set \(R_b\), and the triggering testcase set \(T_b \subseteq R_b\).
These artifacts are then used to instantiate the task inputs for the three APR tasks.
Figure~\ref{fig:testcase-method-dist} shows the distributions of relevant test methods and triggering test methods across the selected bugs.

\subsection{Implementation}
\label{sec:experiment:implementation}

\begin{table}[t]
\centering
\scriptsize
\renewcommand{\arraystretch}{1.2}
\caption{Population size and sampled cases for human annotation.}
\label{tab:human-annotation-sampling}
\begin{tabular}{l|l|c|c}
\hline
\textbf{Task} & \textbf{Model} & \textbf{Population Size \(N\)} & \textbf{Sample Size \(n\)} \\
\hline

\multirow{3}{*}{Triggering Testcase Identification}
& Claude & 1716 & 94 \\
& DeepSeek & 1716 & 94 \\
& GPT-5 & 1716 & 94 \\

\hline

\multirow{3}{*}{Line Coverage Prediction}
& Claude & 832 & 87 \\
& DeepSeek & 832 & 87 \\
& GPT-5 & 832 & 87 \\

\hline

\multirow{3}{*}{Additional Testcase Generation}
& Claude & 2102 & 92 \\
& DeepSeek & 1369 & 90 \\
& GPT-5 & 591 & 87 \\

\hline
\end{tabular}
\end{table}

We implement an automated pipeline to construct task inputs, query models, parse model outputs, and organize the generated artifacts for automatic evaluation and human annotation.
Since the three APR tasks require different inputs and produce different types of outputs, their implementations differ in several task-specific details.
We describe these implementation details below. The exact prompts for all settings are provided in our replication package.

\subsubsection{Triggering Testcase Identification}
\label{sec:implementation-trigger-identification}

For triggering testcase identification, each input contains the buggy functions and relevant testcase methods.
Because some bugs have too many relevant test methods to fit into one context window, we split the relevant methods of each bug into slices of at most 300 methods.
Each slice contains the same buggy functions and a different subset of relevant test methods, and is treated as an independent task instance.
The model is asked to identify the triggering testcases within the slice and generate a repair patch.

\subsubsection{Line Coverage Prediction}

For line coverage prediction, each input contains the buggy functions and all triggering testcases.
Since the number of triggering testcases is usually small, no slicing is needed, and each bug corresponds to one task instance.
The model is asked to predict the execution lines in the buggy functions, generate a repair patch, and predict the execution lines in the model-patched functions.

\subsubsection{Additional Testcase Generation}

For additional testcase generation, each input contains the buggy functions and all triggering testcases.
As in line coverage prediction, no slicing is needed.
The model is asked to generate a repair patch and one or more additional testcases.
We treat each generated additional testcase as an independent instance; multiple testcases generated for the same bug share the same model-generated patch but are evaluated and annotated independently.
If a response contains no additional testcase, it does not contribute an additional-testcase instance, but its generated patch is still retained for patch evaluation.

\subsubsection{Baseline Repair}

We also include a baseline repair setting that does not require task-specific intermediate artifacts.
For each bug, the model receives only the complete developer-modified functions extracted from the buggy program and is asked to produce a minimal correct fix.
Each generated patch is applied to the buggy program and evaluated using the same developer-written test suite and patch-evaluation procedure as in the three task-specific settings.

\subsubsection{Model Querying and Execution Environment}

We query all studied models through their corresponding provider APIs in a zero-shot setting, without fine-tuning or external tool use.
For each task instance, we collect one response from each model.
We use the most deterministic decoding configuration supported by each provider: temperature 0 for Claude Sonnet 4.5, and the provider-default configurations for GPT-5 and DeepSeek-R1 when effective temperature control is unavailable.

For execution-based evaluation, we run all builds, tests, generated testcases, and coverage collection in isolated execution environments.
We use JaCoCo~\citep{jacoco} to collect line-level coverage and map the results back to the extracted buggy or model-patched functions.
Each build, test, and coverage execution is given a timeout of 300 seconds, and timed-out executions are treated as failed executions.
We configure the Java version and build tool according to each \textsc{Defects4J} project; the main build tools include Maven, Gradle, and Ant.
Additional API and environment details are provided in our replication package.

\subsubsection{Population and Sample Size}
\label{sec:experiment:population-sample}

Table~\ref{tab:human-annotation-sampling} reports the population size and sample size for human annotation.
The population size \(N\) is defined according to the instance construction strategy of each task: sliced instances for triggering testcase identification, bug-level instances for line coverage prediction, and generated additional testcase instances for additional testcase generation.
Following the sampling strategy described in Section~\ref{sec:human-annotation}, we randomly sample cases from each task-model group for manual annotation.

\section{Results}
\label{sec:results}
\subsection{Prevalence of Repair and Understanding Hallucinations}
\label{sec:result_RQ1}
In this study, we consider hallucinations at two levels: hallucinations in the final repair artifact and hallucinations in intermediate APR subtask artifacts. 
The intermediate subtasks include triggering testcase identification, line coverage prediction, and additional testcase generation. 

\subsubsection{Repair Hallucination}

\begin{table*}[t]
\scriptsize
\centering
\setlength{\tabcolsep}{3pt}
\renewcommand{\arraystretch}{1.15}
\caption{Number of plausible patches across the baseline setting and three tasks.}
\label{tab:pass_original_testcases}
\begin{tabular}{l|cc|cc|cc|cc}
\toprule
\multirow{2}{*}{Model} 
& \multicolumn{2}{c|}{Baseline}
& \multicolumn{2}{c|}{\makecell{Triggering Testcase\\Identification}}
& \multicolumn{2}{c|}{\makecell{Line Coverage\\Prediction}}
& \multicolumn{2}{c}{\makecell{Additional Testcase\\Generation}} \\
\cmidrule(lr){2-3} 
\cmidrule(lr){4-5} 
\cmidrule(lr){6-7} 
\cmidrule(lr){8-9}
& \Pass & \Fail 
& \Pass & \Fail 
& \Pass & \Fail
& \Pass & \Fail \\
\midrule
Claude   
& 178 (21.4\%) & 654
& 222 (26.7\%) & 610
& 308 (37.0\%) & 524
& 293 (35.2\%) & 539 \\

DeepSeek 
& 175 (21.0\%) & 657
& 314 (37.7\%) & 518
& 403 (48.4\%) & 429
& 310 (37.3\%) & 522 \\

GPT-5    
& 219 (26.3\%) & 613
& 378 (45.4\%) & 454
& 465 (55.9\%) & 367
& 387 (46.5\%) & 445 \\
\bottomrule
\end{tabular}
\end{table*}

\begin{figure*}[t]
    \centering
    \includegraphics[width=\textwidth]{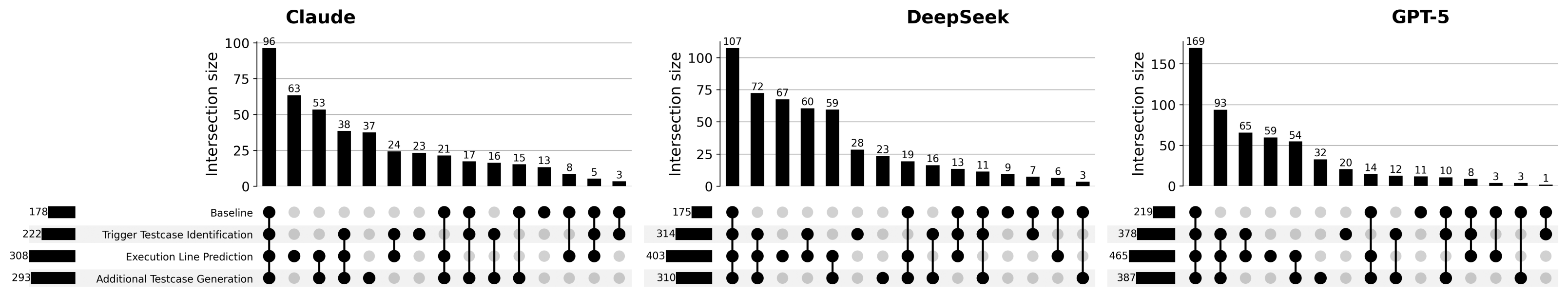}
    \caption{Overlap of plausible patches across the baseline setting and the three task settings.}
    \label{fig:upset_pass}
\end{figure*}

Table~\ref{tab:pass_original_testcases} reports the number of plausible patches generated by each model across the baseline setting and the three tasks. 
A patch is labeled \(\Pass\) if it passes all developer-written testcases; otherwise, it is labeled \(\Fail\), including patches that fail at least one developer-written testcase, patches that cannot be compiled, and cases where executing the developer-written testcases on the patched program results in a timeout.
For triggering testcase identification, relevant testcases may be divided into multiple subsets and queried separately. 
We therefore aggregate results at the bug level: a bug is labeled \Pass only if every subset containing at least one triggering testcase produces a plausible patch; otherwise, it is labeled \Fail. 
The same criterion is used when grouping \Pass and \Fail cases in Figure~\ref{fig:upset_pass}, Figure~\ref{fig:trigger_pred_score}, and Figure~\ref{fig:task3_testcase_validity}.

From Table~\ref{tab:pass_original_testcases}, we observe that plausible patches account for only 21.0\%--55.9\% of repair attempts across models and settings. 
The baseline rates are particularly low, ranging from 21.0\% to 26.3\%. 
These results indicate that repair hallucination remains prevalent in LLM-based APR.

Among the three tasks, line coverage prediction achieves the highest plausible-patch rate for every model, reaching 55.9\% for GPT-5, 48.4\% for DeepSeek, and 37.0\% for Claude. 
This suggests that providing triggering testcases and requiring the model to reason about executed lines may help the model achieve a higher plausible-patch generation rate, which may indicate fewer hallucination-related failures during patch generation.

Across models, GPT-5 consistently produces the most plausible patches in all settings, increasing from 219 in the baseline to 378--465 across the three tasks. 
DeepSeek similarly increases from 175 to 310--403 plausible patches, while Claude also improves over its baseline of 178, although the improvement is much smaller for triggering testcase identification.

Figure~\ref{fig:upset_pass} shows the overlap among bugs plausibly repaired under the four settings. 
The horizontal bars show the total number of plausible patches generated under each setting, while the vertical bars show the number of bugs shared by a particular combination of settings; the filled dots below indicate which settings are included in that intersection.

For each model, the largest intersection corresponds to bugs for which all four settings generate plausible patches.
This intersection contains 96 bugs for Claude, 107 bugs for DeepSeek, and 169 bugs for GPT-5.
This suggests that there exists a core set of bugs for which models can consistently generate plausible patches across different prompting settings. 
However, the remaining intersections are still substantial, indicating that many bugs are plausibly patched only under specific settings or combinations of settings. 
Therefore, the models exhibit limited robustness across APR task settings, as their repair outcomes vary with the specific task formulation and input evidence provided in the prompt.

\rqbox{\textbf{Finding 1.} Repair hallucination remains prevalent in LLM-based APR, as plausible patches account for only a limited portion of all repair attempts across models and settings. GPT-5 produces the largest number of plausible patches, and line coverage prediction yields the highest plausible-patch rate among the three task settings.}

\subsubsection{Triggering Testcase Identification}

\begin{table}[t]
\centering
\caption{Precision, recall, and F1 scores for triggering testcase identification.}
\label{tab:trigger_prf_by_status}
\begin{tabular}{llcccc}
\toprule
Model & Outcome & \#Bugs & Precision & Recall & F1 \\
\midrule
\multirow{2}{*}{Claude}
    & \Pass & 222 & 0.478 & 0.564 & 0.482 \\
    & \Fail & 610 & 0.194 & 0.273 & 0.195 \\
\midrule
\multirow{2}{*}{DeepSeek}
    & \Pass & 314 & 0.444 & 0.502 & 0.445 \\
    & \Fail & 518 & 0.178 & 0.245 & 0.180 \\
\midrule
\multirow{2}{*}{GPT-5}
    & \Pass & 378 & 0.619 & 0.635 & 0.607 \\
    & \Fail & 454 & 0.232 & 0.240 & 0.214 \\
\bottomrule
\end{tabular}
\end{table}

\begin{figure*}[t]
    \centering
    \includegraphics[width=0.95\textwidth]{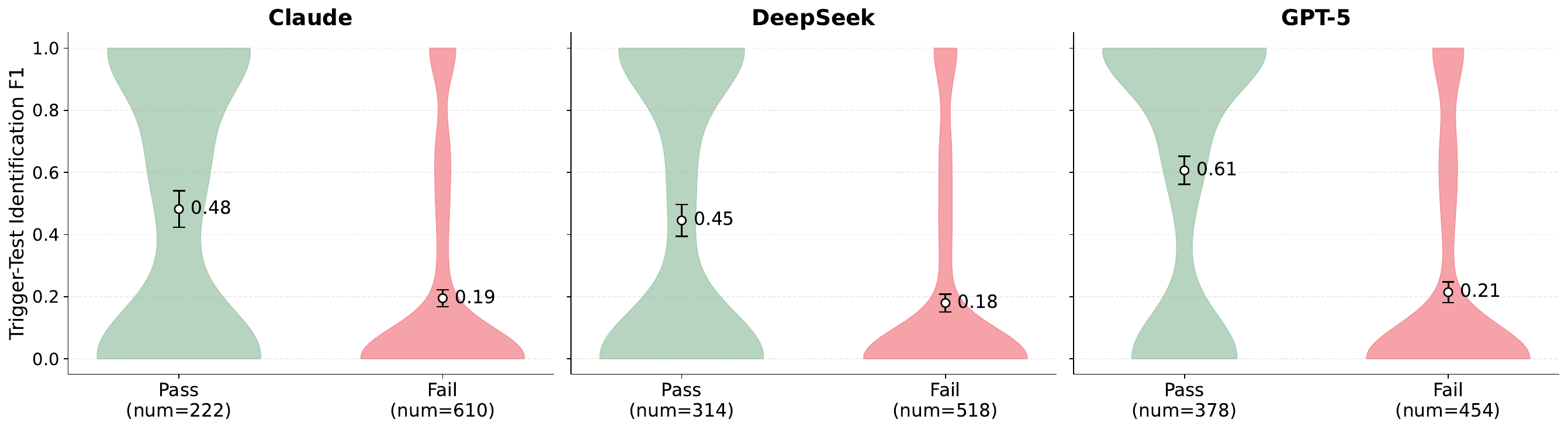}
    \caption{F1-score distributions for the triggering testcase identification task. Dots indicate mean F1 scores, and error bars indicate 95\% confidence intervals.}
    \label{fig:trigger_pred_score}
\end{figure*}

Table~\ref{tab:trigger_prf_by_status} reports the average precision, recall, and F1 for triggering testcase identification across bugs, while Figure~\ref{fig:trigger_pred_score} shows the corresponding distributions of F1 scores for individual bugs.
In each subfigure, the dots indicate the average prediction success rates, and the error bars denote 95\% confidence intervals.  

Across all three models, passing repairs consistently achieve substantially higher identification performance than failed repairs. For example, Claude obtains an F1 score of 0.482 for passing repairs, compared with 0.195 for failed repairs, while DeepSeek exhibits a similar gap, with scores of 0.445 and 0.180, respectively. GPT-5 achieves the strongest overall performance, attaining an F1 score of 0.607 for passing repairs versus 0.214 for failed repairs.

The distributions in Figure~\ref{fig:trigger_pred_score} further suggest that bugs that are successfully repaired are more likely to involve correct identification of bug-triggering testcases. However, the substantial overlap between the distributions indicates that this relationship is not absolute, as some failed repairs achieve high F1 scores while some passing repairs receive low scores.

Overall, these results show a clear association between triggering-testcase identification and repair success. On average, passing repairs achieve higher precision, recall, and F1 scores across all three models, indicating that successful repairs are less likely to involve understanding hallucinations at this stage. Nevertheless, the overlap between passing and failed repairs suggests that correct testcase identification alone does not fully determine whether a repair will succeed.

\subsubsection{Line Coverage Prediction}


\begin{table*}[t]
\centering
\scriptsize
\setlength{\tabcolsep}{2.5pt}
\renewcommand{\arraystretch}{1.05}
\caption{Precision (P), recall (R), and F1 scores for line coverage prediction on the buggy and model-patched programs.}
\label{tab:execline_prf}
\begin{tabular}{@{}llc|ccc|ccc@{}}
\toprule
\multirow{2}{*}{Model}
& \multirow{2}{*}{Outcome}
& \multirow{2}{*}{\# Bugs}
& \multicolumn{3}{c|}{Buggy}
& \multicolumn{3}{c}{Model-Patched} \\
\cmidrule(lr){4-6}
\cmidrule(lr){7-9}
& & & P & R & F1 & P & R & F1 \\
\midrule

\multirow{3}{*}{Claude}
& \Pass
& 308
& 0.706 & 0.881 & 0.751
& 0.745 & 0.885 & 0.779 \\

& \Notpass
& 416
& 0.664 & 0.860 & 0.714
& 0.647 & 0.845 & 0.697 \\

& \UncompilableTimeout
& 108
& 0.671 & 0.864 & 0.716
& -- & -- & -- \\

\cmidrule(lr){1-9}

\multirow{3}{*}{DeepSeek}
& \Pass
& 403
& 0.888 & 0.881 & 0.866
& 0.955 & 0.911 & 0.922 \\

& \Notpass
& 296
& 0.783 & 0.799 & 0.767
& 0.757 & 0.809 & 0.757 \\

& \UncompilableTimeout
& 133
& 0.845 & 0.837 & 0.816
& -- & -- & -- \\

\cmidrule(lr){1-9}

\multirow{3}{*}{GPT-5}
& \Pass
& 465
& 0.897 & 0.883 & 0.872
& 0.741 & 0.692 & 0.707 \\

& \Notpass
& 291
& 0.791 & 0.783 & 0.753
& 0.492 & 0.506 & 0.482 \\

& \UncompilableTimeout
& 76
& 0.831 & 0.794 & 0.795
& -- & -- & -- \\

\bottomrule
\end{tabular}
\end{table*}

\begin{figure*}[t]
    \centering
    \begin{subfigure}{0.95\textwidth}
        \centering
        \includegraphics[width=\linewidth]{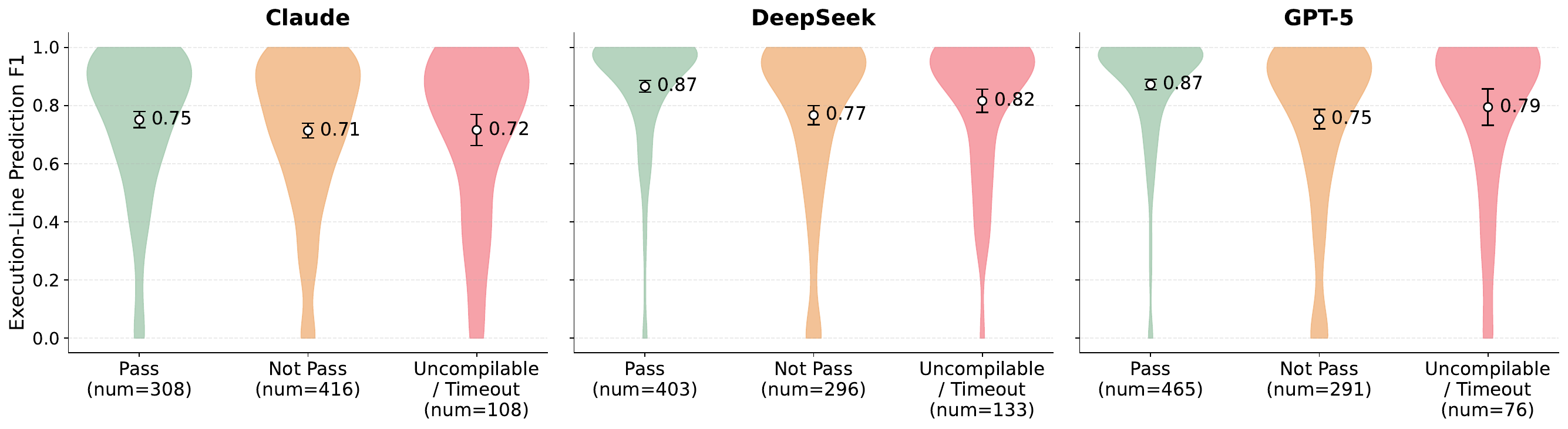}
        \caption{Line coverage prediction on the buggy program}
        \label{fig:pre_repair_line_pred_score}
    \end{subfigure}
    \hfill
    \begin{subfigure}{0.95\textwidth}
        \centering
        \includegraphics[width=\linewidth]{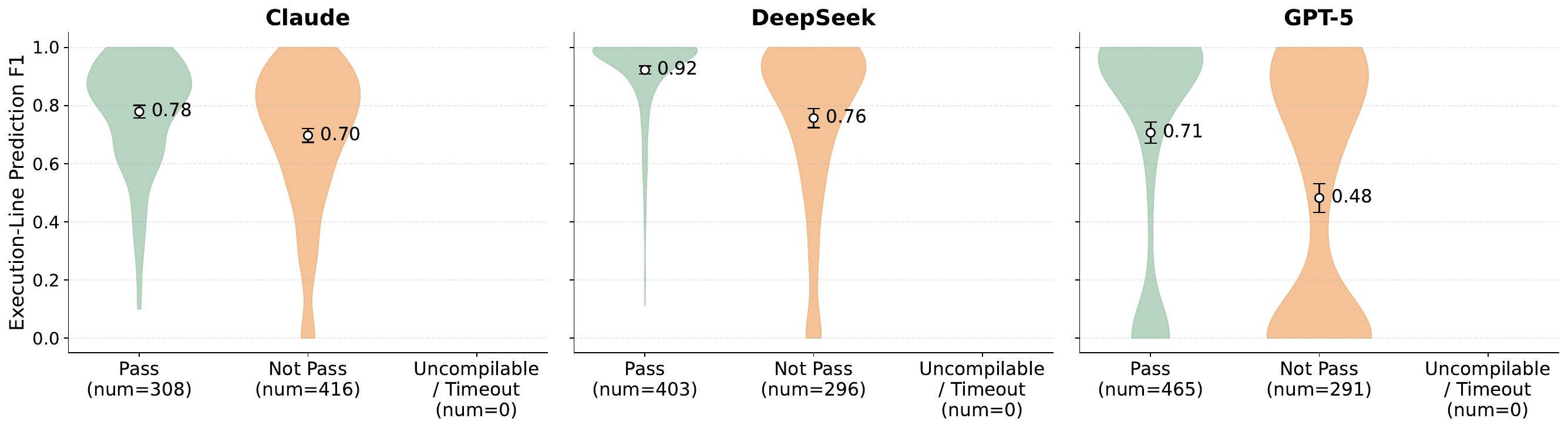}
        \caption{Line coverage prediction on the model-patched program}
        \label{fig:post_repair_line_pred_score}
    \end{subfigure}
    \caption{F1-score distributions for line coverage prediction on the buggy and model-patched programs. Dots indicate mean F1 scores, and error bars indicate 95\% confidence intervals.}
    \label{fig:line_pred_score}
\end{figure*}

This task evaluates whether the model can predict the lines executed by the given triggering testcases before and after applying its generated patch. 
Table~\ref{tab:execline_prf} reports the precision, recall, and F1 scores for line coverage prediction on the buggy and model-patched programs, while Figure~\ref{fig:line_pred_score} shows the corresponding per-bug F1-score distributions. 
We report \UncompilableTimeout separately from \Notpass because our instrumentation tool cannot collect line coverage from model-patched programs that fail to compile or execute within the timeout.

Across all three models, the \Pass group achieves higher mean F1 scores on the buggy program than both the \Notpass and \UncompilableTimeout groups. 
Specifically, the \Pass scores reach 0.751 for Claude, 0.866 for DeepSeek, and 0.872 for GPT-5, compared with 0.714--0.816 across the two unsuccessful outcome groups. 
The same pattern holds on the model-patched program, where the \Pass group consistently outperforms the \Notpass group. 
This result suggests that, on average, plausible patch generation is associated with a more faithful understanding of the execution behavior of both the original buggy code and the model-generated patch, indicating fewer execution-behavior hallucinations.

\subsubsection{Additional Testcase Generation}

\begin{figure*}[t]
    \centering
    \includegraphics[width=0.95\textwidth]{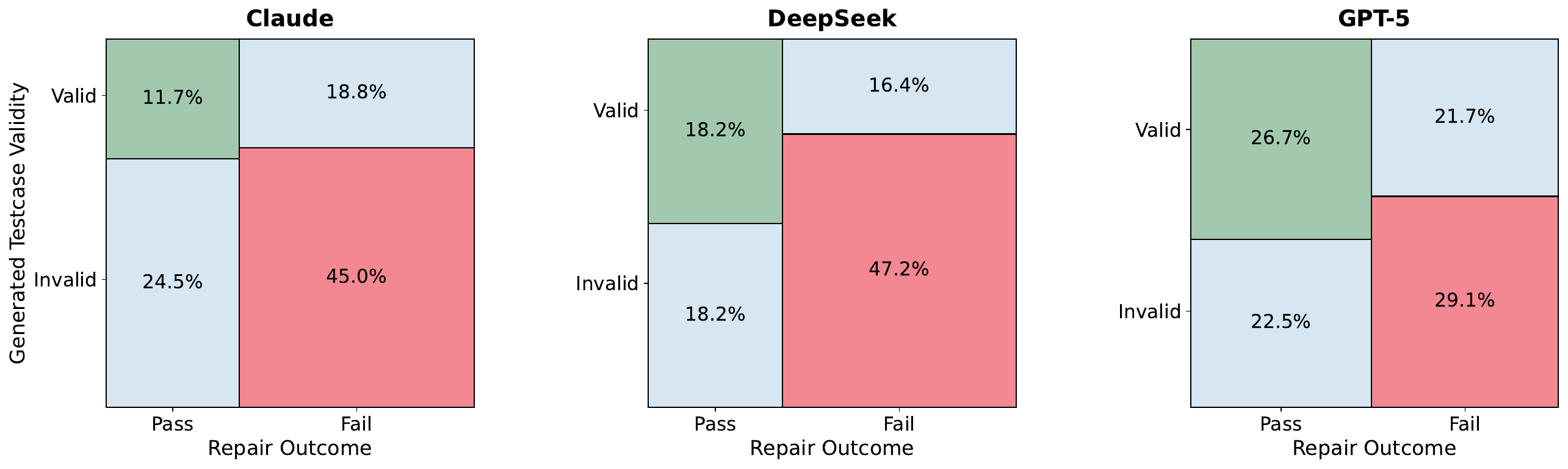}
    \caption{Distributions of generated triggering testcase validity and repair outcome across models. Rectangle areas are proportional to the corresponding percentages among all generated testcases for each model.}
    \label{fig:task3_testcase_validity}
\end{figure*}

Figure~\ref{fig:task3_testcase_validity} shows the distribution of generated triggering testcase validity and repair outcome for each model. 
A generated testcase is valid if it fails on the buggy program and passes on the developer-written fixed program. 
Percentages are normalized over all generated additional testcases for each model, with multiple testcases from the same task instance counted independently.

Overall, generating valid triggering testcases remains challenging. 
Valid testcases account for only 30.5\% of Claude's outputs, 34.6\% of DeepSeek's outputs, and 48.4\% of GPT-5's outputs, meaning that more than half of the generated testcases are invalid for all three models.

Across all three models, \Pass repairs have a higher proportion of valid generated triggering testcases than \Fail repairs. 
For example, for DeepSeek, 50.0\% of the generated triggering testcases associated with \Pass repairs are valid, compared with 25.8\% for \Fail repairs.
However, the validity of generated triggering testcases is not fully aligned with repair success. 
Across models, 45.7\%--67.7\% of the generated triggering testcases associated with \Pass repairs are still invalid, while 25.8\%--42.7\% of those associated with \Fail repairs are valid. 
Thus, passing the developer-written test suite does not necessarily imply a faithful understanding of the bug behavior, while generating a valid triggering testcase does not guarantee a plausible repair.

Overall, the three intermediate APR tasks reveal a consistent pattern.
Understanding hallucinations in intermediate artifacts still account for a considerable portion of model outputs, indicating that hallucination is not limited to the final repair artifact.
At the same time, \Pass cases generally produce more accurate intermediate artifacts than \Fail cases.
One plausible explanation is that successful repair may require the model to construct a reasonably accurate representation of the bug before producing the patch.
Therefore, \Pass cases tend to exhibit fewer understanding hallucinations than \Fail cases across the three tasks.

However, this association is not deterministic.
Some \Fail cases still contain accurate intermediate artifacts, suggesting that understanding certain aspects of the bug may be insufficient to produce a plausible patch.
Conversely, some \Pass cases are accompanied by inaccurate or invalid intermediate artifacts, showing that passing the developer-written test suite does not necessarily imply a fully faithful understanding of the bug behavior.

Across models, GPT-5 generally produces the most accurate intermediate artifacts, followed by DeepSeek and Claude. 
Nevertheless, understanding hallucinations remain evident across all three models, including the strongest-performing model. 
To further characterize this problem, the next RQ examines the concrete types and distributions of understanding hallucinations across APR tasks and models.




\rqbox{\textbf{Finding 2.} Hallucinations in intermediate understanding artifacts remain common in LLM-based APR. Plausible repairs are generally associated with more accurate understanding artifacts, but this relationship is not deterministic. Among the evaluated models, GPT-5 shows the lowest tendency toward understanding hallucination across the three intermediate tasks.}

\subsection{Hallucination Types and Distributions}
\label{sec:result_RQ2}

In this section, we examine how hallucinations manifest in final repair artifacts and intermediate APR artifacts, and develop a taxonomy of observed hallucination types.

\subsubsection{Repair Hallucination}
\begin{table*}[t]
\centering
\scriptsize
\caption{Distribution of repair hallucination categories across the three APR artifact-generation tasks.
Each cell reports counts in the order of Claude / DeepSeek / GPT-5.}
\label{tab:repair-hallucination-classification}

\renewcommand{\arraystretch}{1.15}
\setlength{\tabcolsep}{2.5pt}

\begin{tabularx}{\textwidth}{
@{}
>{\raggedright\arraybackslash}p{4.4cm}|
>{\centering\arraybackslash}p{1.7cm}|
*{3}{>{\centering\arraybackslash}X}
@{}
}
\hline
\multirow{2}{*}{\textbf{Category}}
& \multirow{2}{*}{\textbf{$T_{\text{patch}}$}}
& \textbf{Triggering Testcase}
& \textbf{Line Coverage}
& \textbf{Additional Testcase} \\
&
& \textbf{Identification}
& \textbf{Prediction}
& \textbf{Generation} \\
\hline

Non-existent Symbol Reference
& \multirow{3}{*}{\Uncompilable}
& 4 / 5 / 6
& 8 / 12 / 0
& 16 / 8 / 6 \\

Missing Import or Dependency
&
& 0 / 3 / 0
& 2 / 1 / 0
& 5 / 0 / 0 \\

Syntax or Structural Error
&
& 1 / 2 / 2
& 1 / 1 / 1
& 0 / 4 / 2 \\

\hline

Incorrect Repair Strategy
& \multirow{6}{*}{\shortstack{\Pass,\\\Notpass}}
& 15 / 13 / 12
& 11 / 11 / 18
& 7 / 12 / 10 \\

Incorrect Causal Localization
&
& 40 / 44 / 36
& 28 / 31 / 23
& 17 / 27 / 25 \\

Incorrect API Usage
&
& 0 / 1 / 1
& 9 / 0 / 0
& 3 / 0 / 1 \\

Partial Semantic Repair
&
& 4 / 4 / 8
& 4 / 5 / 5
& 6 / 6 / 10 \\

Extraneous Conditional Logic
&
& 0 / 1 / 2
& 1 / 1 / 1
& 4 / 2 / 2 \\

Incorrect Boundary Check
&
& 2 / 1 / 2
& 1 / 2 / 5
& 2 / 3 / 2 \\

\hline

Test-Specific Heuristic Overfitting
& \Pass
& 0 / 0 / 1
& 1 / 1 / 2
& 4 / 1 / 2 \\

\hline

Others
& \All
& 0 / 0 / 0
& 2 / 0 / 0
& 3 / 1 / 1 \\

\hline

\textbf{Sample Size}
& \textbf{--}
& \textbf{94 / 94 / 94}
& \textbf{87 / 87 / 87}
& \textbf{92 / 90 / 87} \\

\hline
\end{tabularx}
\end{table*}

Following the human annotation procedure described in Section~\ref{sec:human-annotation}, we categorize the observed repair hallucinations into 10 types based on their manifestations in generated patches, together with an \emph{Others} category that groups infrequent types observed only once.
Table~\ref{tab:repair-hallucination-classification} orders these categories from those most directly observable through
execution results to those less directly observable from execution results alone.
The \textit{Category} column lists the hallucination type assigned during labeling, and the $T_{\text{patch}}$ column indicates the execution results in which the hallucination may appear: \Pass, \Notpass, or \Uncompilable.
For each APR task, the table reports the number of observed cases in the order of Claude / DeepSeek / GPT-5, followed by the corresponding sample sizes.

Specifically, we first present hallucinations that make the generated patch uncompilable, followed by hallucinations that lead to compilable but test-failing patches, hallucinations that may appear in both passing and failing patches, and finally hallucinations that pass the developer-written tests but require semantic inspection to identify.
Following prior APR literature on patch overfitting~\cite{smith2015cure,xin2017identifying}, we treat hallucinated patches that pass all developer-written test cases as overfitting patches, because they satisfy the available test suite but do not actually repair the bug and deviate from the developer-intended repair semantics.

To further clarify the taxonomy, we define each hallucination type below. 
The first three categories capture hallucinations that make the generated patch break project compilation. 
These errors are the most directly observable because they prevent the repaired project from being built successfully.

$\bullet$~\textbf{Non-existent Symbol Reference} refers to a repair hallucination in which the generated patch references a method, type, variable, field, class, or other program symbol that does not exist in the current codebase or accessible dependencies.

$\bullet$~\textbf{Missing Import or Dependency} refers to a repair hallucination in which the generated patch uses valid program symbols or external libraries whose required import statements or dependency declarations are absent from the current file or project context, causing them to be unresolved during compilation.

$\bullet$~\textbf{Syntax or Structural Error} refers to a repair hallucination in which the generated patch violates the syntactic or structural integrity of the program, such as by introducing syntactically invalid statements, unmatched braces, misplaced declarations, or incorrectly nested code blocks.
For example, in the patch generated by Claude for Closure-138 in the line coverage prediction task, the model introduces an extra closing brace that breaks the surrounding method structure, causing compilation errors.

The next six categories produce compilable patches and can manifest as either overfitting \Pass patches or \Notpass patches, depending on whether the developer-written test cases expose the hallucinated repair behavior.
Unlike hallucinations involving compilation errors, these cases become observable only through test execution or semantic inspection.
When the hallucinated behavior is exercised by the test suite, the patch is classified as \Notpass; otherwise, it may pass all developer-written tests despite deviating from the intended repair semantics.
We identify such overfitting patches through manual semantic comparison with the developer-written repair, which serves as the reference for the intended repair semantics.

$\bullet$~\textbf{Incorrect Repair Strategy} refers to a repair hallucination in which the generated patch targets fault-relevant statements in the buggy function but follows a repair strategy that fundamentally deviates from the developer-intended fix. 

$\bullet$~\textbf{Incorrect Causal Localization} refers to a repair hallucination in which the generated patch modifies a non-causal part of the program, rather than addressing the actual root cause of the bug required by the developer-intended repair.

\begin{figure}[t]
\centering
\begin{subfigure}[t]{0.48\textwidth}
\centering
\includegraphics[width=\linewidth]{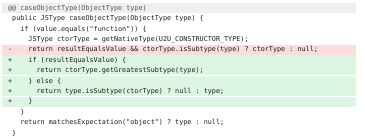}
\caption{Developer diff}
\label{fig:closure7-developer-diff}
\end{subfigure}
\hfill
\begin{subfigure}[t]{0.48\textwidth}
\centering
\includegraphics[width=\linewidth]{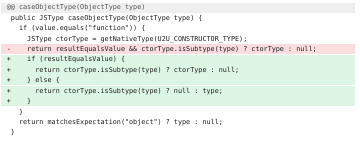}
\caption{Model-generated diff}
\label{fig:closure7-model-diff}
\end{subfigure}

\caption{Example of incorrect API usage in the patch generated by Claude for Closure-7 in the additional testcase generation task.}
\label{fig:incorrect-api-closure7}
\end{figure}

$\bullet$~\textbf{Incorrect API Usage} refers to a repair hallucination in which the generated patch modifies the relevant program location and broadly follows the intended repair direction, but uses an API in a semantically incorrect way, such as by calling an inappropriate API, passing incorrect arguments, or misinterpreting the API's return semantics.
For example, Figure~\ref{fig:incorrect-api-closure7} shows the developer-written and model-generated patches for Closure-7 in the additional testcase generation task.
The developer fix uses \sccode{ctorType.getGreatestSubtype(type)} to compute the refined function subtype when \sccode{resultEqualsValue} is true, and uses \sccode{type.isSubtype(ctorType)} in the other branch.
In contrast, the model-generated patch continues to use \sccode{ctorType.isSubtype(type)} in both branches.
Although the patch modifies the correct conditional logic, it uses the API incorrectly by preserving the wrong subtype relation, making it an incorrect API usage hallucination.

\begin{figure}[t]
\centering

\begin{subfigure}[t]{0.48\textwidth}
\centering
\includegraphics[width=\linewidth]{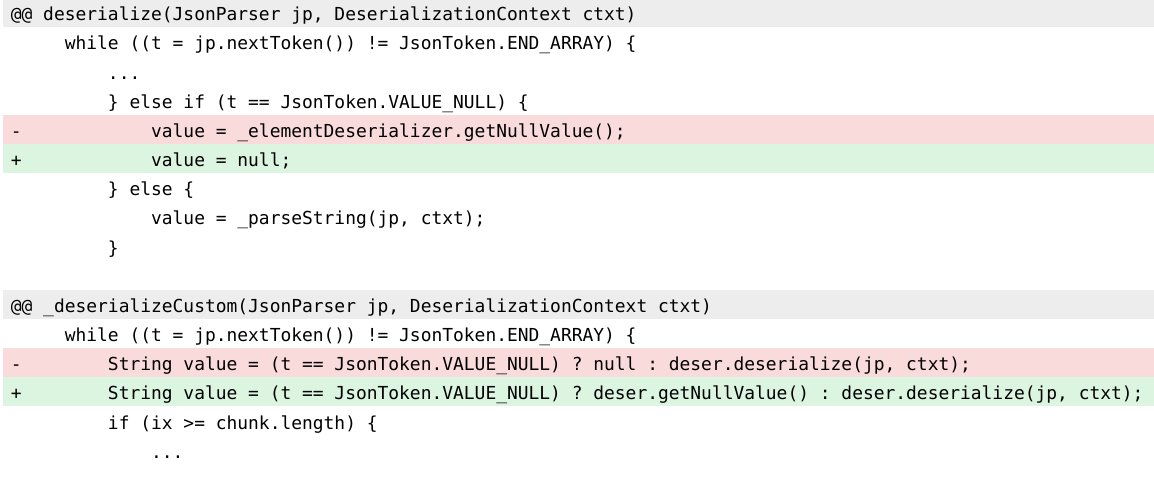}
\caption{Developer diff}
\label{fig:jacksondatabind3-developer-diff}
\end{subfigure}
\hfill
\begin{subfigure}[t]{0.48\textwidth}
\centering
\includegraphics[width=\linewidth]{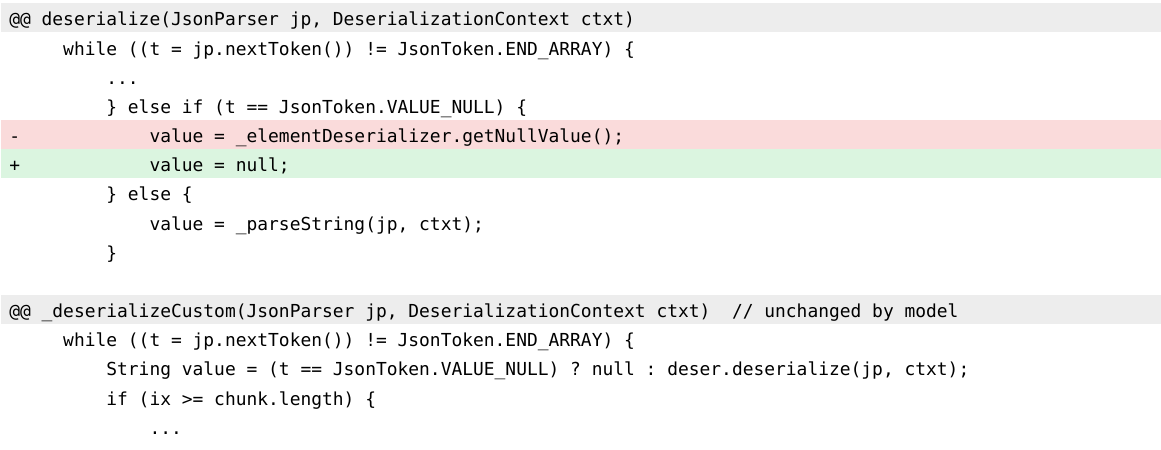}
\caption{Model-generated diff}
\label{fig:jacksondatabind3-model-diff}
\end{subfigure}

\caption{Example of partial semantic repair in the patch generated by Claude for JacksonDatabind-3 in the line coverage prediction task.}
\label{fig:partial-semantic-jacksondatabind3}
\end{figure}

$\bullet$~\textbf{Partial Semantic Repair} refers to a repair hallucination in which the bug requires coordinated changes across multiple locations, cases, or execution paths, but the generated patch implements only a subset of the developer-intended fix.
For example, Figure~\ref{fig:partial-semantic-jacksondatabind3} illustrates an instance of partial semantic repair in the patch generated by Claude for JacksonDatabind-3 in the line coverage prediction task.
The developer fix updates two null-handling paths, whereas the model-generated patch fixes only one of them.

\begin{figure}[t]
\centering
\begin{subfigure}[t]{0.48\textwidth}
\centering
\includegraphics[width=\linewidth]{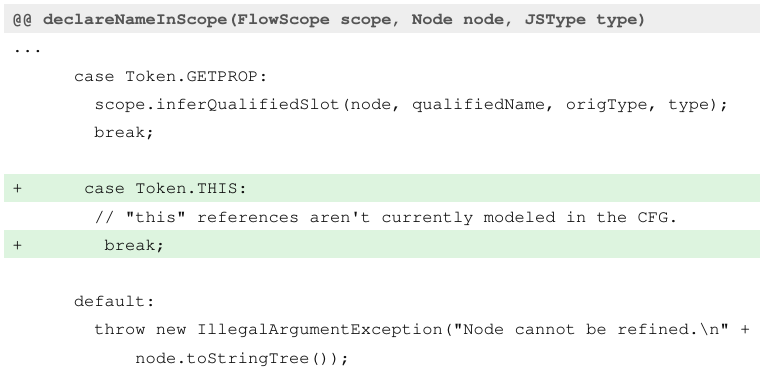}
\caption{Developer diff}
\label{fig:closure19-developer-diff}
\end{subfigure}
\hfill
\begin{subfigure}[t]{0.48\textwidth}
\centering
\includegraphics[width=\linewidth]{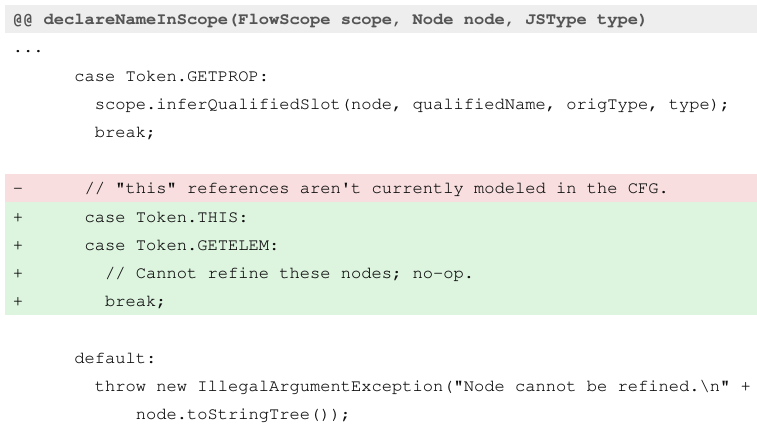}
\caption{Model-generated diff}
\label{fig:closure19-model-diff}
\end{subfigure}

\caption{Example of extraneous conditional logic in the patch generated by GPT-5 for Closure-19 in the triggering testcase identification task.}
\label{fig:extraneous-conditional-closure19}
\end{figure}

$\bullet$~\textbf{Extraneous Conditional Logic} refers to a repair hallucination in which the generated patch modifies the relevant program location and broadly follows the intended repair direction, but introduces additional conditional checks, guard branches, or special-case handling that is not required by the developer-intended fix.
For example, Figure~\ref{fig:extraneous-conditional-closure19} illustrates an instance of extraneous conditional logic in the patch generated by GPT-5 for Closure-19 in the triggering testcase identification task.
The developer fix adds a no-op case only for \sccode{Token.THIS}.
In contrast, the model-generated patch additionally treats \sccode{Token.GETELEM} as a no-op case.
Thus, the model introduces extra conditional logic beyond the developer-intended fix, making it an extraneous conditional logic hallucination.

\begin{figure}[t]
\centering
\begin{subfigure}[t]{0.48\textwidth}
\centering
\includegraphics[width=\linewidth]{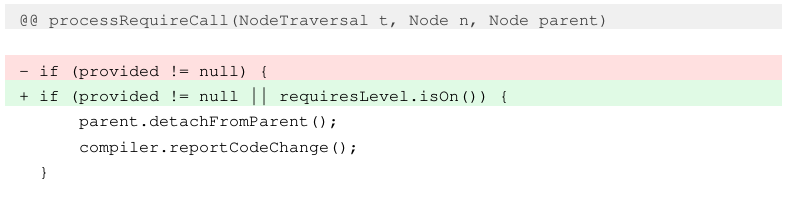}
\caption{Developer diff}
\label{fig:closure113-developer-fix}
\end{subfigure}
\hfill
\begin{subfigure}[t]{0.48\textwidth}
\centering
\includegraphics[width=\linewidth]{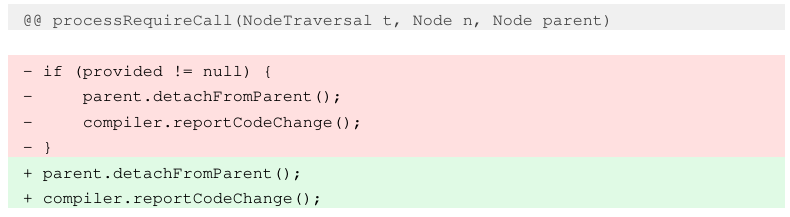}
\caption{Model-generated diff}
\label{fig:closure113-model-fix}
\end{subfigure}

\caption{Example of incorrect boundary check in the patch generated by Claude for Closure-113 in the line coverage prediction task.}
\label{fig:incorrect-boundary-closure113}
\end{figure}

$\bullet$~\textbf{Incorrect Boundary Check} refers to a repair hallucination in which the generated patch targets the correct program location and broadly follows the intended repair direction, but encodes an incorrect condition, such as an overly broad or overly restrictive guard, an incorrect predicate, or a wrong threshold value.
For example, Figure~\ref{fig:incorrect-boundary-closure113} illustrates an instance of incorrect boundary check in the patch generated by Claude for Closure-113 in the line coverage prediction task.
The developer fix broadens the guard from \sccode{provided != null} to \sccode{provided != null || requiresLevel.isOn()}, whereas the model-generated patch removes the guard entirely.
As a result, the model introduces an overly broad condition for applying the require-removal logic, making it an incorrect boundary check hallucination.

Finally, we consider hallucinations that pass all developer-written test cases but remain semantically incorrect. These cases are not exposed by test execution alone and require manual inspection against the developer-intended repair semantics.

\begin{figure}[t]
\centering
\begin{subfigure}[t]{0.48\textwidth}
\centering
\includegraphics[width=\linewidth]{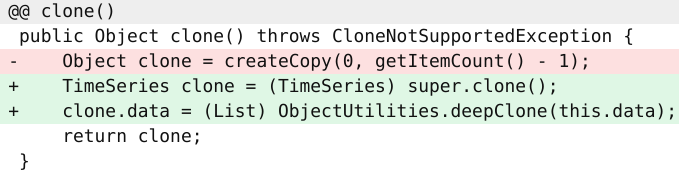}
\caption{Developer diff}
\label{fig:chart17-developer-diff}
\end{subfigure}
\hfill
\begin{subfigure}[t]{0.48\textwidth}
\centering
\includegraphics[width=\linewidth]{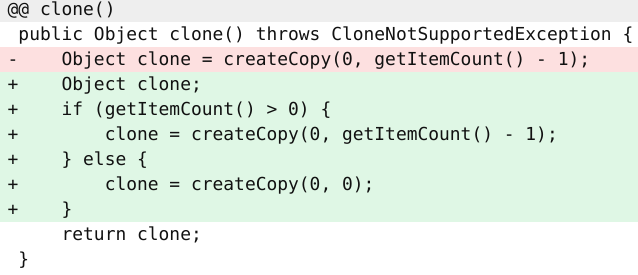}
\caption{Model-generated diff}
\label{fig:chart17-model-diff}
\end{subfigure}
\caption{Example of test-specific heuristic overfitting in the patch generated by Claude for Chart-17 in the additional testcase generation task.}
\label{fig:test-specific-chart17}
\end{figure}

$\bullet$~\textbf{Test-Specific Heuristic Overfitting} refers to a repair hallucination in which the generated patch passes all developer-written testcases by introducing a narrow heuristic tailored to the observed failing behavior. Unlike incorrect repair strategy hallucinations, which generally fail the developer-written testcases due to a fundamentally wrong repair direction, this type can still pass the available tests despite deviating from the intended repair semantics.
For example, Figure~\ref{fig:test-specific-chart17} illustrates an instance of test-specific heuristic overfitting in the patch generated by Claude for Chart-17 in the additional testcase generation task.
The developer fix implements cloning by calling \sccode{super.clone()} and deep-cloning the internal data list.
In contrast, the model-generated patch only adds a special case for empty series and still relies on \sccode{createCopy}.
This heuristic may avoid the observed failure caused by \sccode{createCopy(0, -1)}, but it does not implement the intended cloning semantics, making it a test-specific heuristic overfitting hallucination.

$\bullet$~\textbf{Others} groups infrequent repair hallucination types that are each observed only once and therefore do not form sufficiently recurring categories in our taxonomy.
Representative examples include \emph{Missing Guard Check}, where the patch omits a necessary \sccode{if} check before an operation and therefore executes it without validating the required precondition, and \emph{Extra New Function}, where the patch introduces an unnecessary new function that conflicts with the existing code or program structure, among other one-off hallucination types.
The complete list of labels and their descriptions is provided in our online replication package.

Table~\ref{tab:repair-hallucination-classification} shows that repair hallucinations are prevalent across the three artifact-generation tasks, affecting 72.7\% of the manually inspected repairs, with similar proportions across tasks.

Incorrect causal localization is the dominant category across tasks and models, followed by incorrect repair strategy. 
This suggests that the main challenge for LLM-based APR is not merely editing the faulty region, but correctly identifying the root cause and translating it into a semantically appropriate repair strategy.
Compilation-breaking hallucinations also remain non-negligible, indicating that LLM-generated repairs may still violate symbol-resolution, syntactic, or dependency constraints required by the target project.

Moreover, overfitting patches account for 16.38\% of the observed repair hallucinations. 
Such patches may involve hallucinations including Incorrect Causal Localization, Incorrect API Usage, Partial Semantic Repair, Extraneous Conditional Logic, Incorrect Boundary Check, and Test-Specific Heuristic Overfitting. 
When developer-written test suites provide insufficient behavioral coverage, these patches may pass all available tests despite deviating from the intended repair semantics and therefore remain undetected by automatic validation.
This finding highlights that the effectiveness of execution-based hallucination detection is fundamentally constrained by the adequacy of the available test suite. 
Future research should therefore complement developer-written tests with additional validation mechanisms.

\rqbox{\textbf{Finding 3.}
Repair hallucinations are dominated by incorrect causal localization and inappropriate repair strategies, highlighting root-cause localization and repair-strategy formulation as major obstacles to reliable LLM-based APR.
Compilation errors also remain non-negligible.
Meanwhile, overfitting patches occur frequently and can escape automatic evaluation when it relies primarily on developer-written testcases.}

\subsubsection{Triggering Testcase Identification}

\begin{table}[t]
\centering
\scriptsize
\caption{Distribution of understanding hallucinations in the Triggering Testcase Identification task. Counts in the \# Bugs column are reported in the order of Claude / DeepSeek / GPT-5.}
\label{tab:task1-trigger-understanding-hallucination}
\renewcommand{\arraystretch}{1.3}
\setlength{\tabcolsep}{4pt}
\begin{tabular}{l|l|c}
\hline
\textbf{Category}
& \textbf{Condition}
& \textbf{\# Bugs} \\
\hline

Correct Identification
& $\hat{T}_b = T_b$
& 191 / 276 / 340 \\

Partial Identification
& $\emptyset \subset \hat{T}_b \subset T_b$
& 40 / 40 / 72 \\

Trigger Omission
& $\hat{T}_b = \emptyset$
& 54 / 71 / 14 \\

Spurious Identification
& $(T_b \subseteq \hat{T}_b) \wedge (\hat{T}_b \setminus T_b \neq \emptyset)$
& 151 / 168 / 102 \\

Mixed Misidentification
& $(T_b \setminus \hat{T}_b \neq \emptyset) \wedge (\hat{T}_b \setminus T_b \neq \emptyset)$
& 396 / 277 / 304 \\

\hline
\textbf{Total}
& --
& 832 / 832 / 832 \\
\hline
\end{tabular}
\end{table}

While RQ1 shows that plausible repairs are associated with higher aggregate F1 scores, it does not reveal the concrete error patterns behind incorrect trigger predictions.
During manual annotation, we found that low prediction success rates can result from different failure patterns.
To characterize how triggering-testcase understanding hallucinations manifest beyond the aggregate F1 results in RQ1, we compare the model-predicted trigger set $\hat{T}_b$ with the oracle trigger set $T_b$ for each bug.
Based on their set relationship, Table~\ref{tab:task1-trigger-understanding-hallucination} classifies each prediction into five identification outcomes, distinguishing exact identification from omissions, spurious additions, and mixed errors.



$\bullet$~\textbf{Correct Identification} refers to the case where the set of triggering testcases identified by the model exactly matches the oracle trigger testcase set.

$\bullet$~\textbf{Partial Identification} refers to the case where the model identifies only a non-empty subset of the true trigger testcases.

$\bullet$~\textbf{Trigger Omission} refers to the case where the model fails to identify any true trigger testcase.

$\bullet$~\textbf{Spurious Identification} refers to the case where the model includes all true trigger testcases but also introduces unrelated testcases.

$\bullet$~\textbf{Mixed Misidentification} refers to the case where the model both omits some true trigger testcases and introduces unrelated testcases.

Table~\ref{tab:task1-trigger-understanding-hallucination} shows that understanding hallucinations are common in the triggering testcase identification task.
Even GPT-5, which achieves the most correct identifications, still misidentifies the trigger set for more than half of the bugs.
This indicates that correctly identifying all and only the true triggers remains challenging.

Among the hallucination categories, mixed misidentification is the dominant failure pattern.
This indicates that models often do not simply omit trigger testcases or add spurious ones in isolation; instead, they frequently omit true triggers and introduce spurious triggers at the same time.
Spurious identification is also frequent, whereas complete trigger omission is less common, especially for GPT-5.
Together, these patterns suggest that models often capture some bug-related signal but struggle to draw a precise boundary between true triggering testcases and non-triggering ones, revealing limited robustness in triggering testcase identification.

Beyond characterizing how accurately models identify triggering testcases, we also examine whether the availability of oracle triggering testcases in the input is associated with repair success.
This complementary analysis explores whether access to trigger-related behavioral evidence is associated with the model's ability to generate a plausible patch.

\paragraph{Is trigger availability associated with repair success?}

\begin{table}[t]
\centering
\footnotesize
\caption{Relationship between trigger testcase artifact availability and repair success in the Triggering Testcase Identification task. Restricted to bugs that have both a trigger-present ($A_b$) and a trigger-absent ($\bar{A}_b$) variant, so that both conditions are actually tested. Counts in the \# Bugs column are reported in the order of Claude / DeepSeek / GPT-5.}
\label{tab:task1-trigger-classification}
\renewcommand{\arraystretch}{1.2}
\begin{tabular}{l|c|c|c}
\hline
\textbf{Category}
& \textbf{$A_b$}
& \textbf{$\bar{A}_b$}
& \textbf{\# Bugs} \\
\hline
Trigger-insensitive success
& \Pass
& \Pass
& 71 / 91 / 113 \\

Trigger-dependent success
& \Pass
& \Fail
& 17 / 47 / 59 \\

Trigger-absent-only success
& \Fail
& \Pass
& 50 / 40 / 36 \\

Universally failed bugs
& \Fail
& \Fail
& 266 / 226 / 196 \\
\hline
\textbf{Total}
& --
& --
& 404 / 404 / 404 \\
\hline
\end{tabular}
\end{table}

In the triggering testcase identification task, some bugs are represented by multiple input slices, as described in Section~\ref{sec:experiment:implementation}.
These slices contain different subsets of relevant testcase methods: some include at least one oracle trigger testcase, whereas others contain no oracle trigger testcase.
To examine whether the availability of a trigger testcase artifact is associated with repair success, we compare split instances with and without such an artifact for the same bug.

For each bug $b$, we define $A_b$ as the trigger-present group, i.e., the set of input slices whose inputs contain at least one oracle trigger testcase, and $\bar{A}_b$ as the trigger-absent group, i.e., the set of input slices whose inputs contain no oracle trigger testcase.
For each group, we assign \(\Pass\) if at least one input slice produces a plausible patch, and \(\Fail\) otherwise.
Based on the outcomes of $A_b$ and $\bar{A}_b$, we classify each bug into four categories, as shown in Table~\ref{tab:task1-trigger-classification}.
This paired comparison requires each bug to have at least one slice in both groups.
If a bug has only trigger-present slices, its repair outcome under the other condition is not observed and therefore cannot be compared.
We therefore restrict the analysis to the 404 bugs for which both $A_b$ and $\bar{A}_b$ are non-empty.

\textit{Trigger-insensitive success} denotes bugs for which both $A_b$ and $\bar{A}_b$ contain at least one plausible patch, indicating that successful repair is observed regardless of whether an oracle trigger testcase is available in the input.
\textit{Trigger-dependent success} denotes bugs for which only $A_b$ contains a plausible patch, meaning that repair success is observed only when the input includes an oracle trigger testcase.
Conversely, \textit{Trigger-absent-only success} denotes bugs for which only $\bar{A}_b$ contains a plausible patch.
Finally, \textit{Universally failed bugs} are those for which neither group contains a plausible patch.

Table~\ref{tab:task1-trigger-classification} shows that Trigger-insensitive success is the most frequent successful-repair pattern across all three models.
Specifically, 71, 91, and 113 bugs can be successfully repaired both with and without an oracle trigger testcase for Claude, DeepSeek, and GPT-5, respectively.
This result indicates that explicit trigger testcase availability is not necessary for many successfully repaired bugs.
Moreover, for DeepSeek and GPT-5, Trigger-dependent success occurs in 47 and 59 bugs, respectively, exceeding the 40 and 36 Trigger-absent-only success cases.
This pattern suggests that oracle trigger testcases provide useful repair information for these two models.
In contrast, Claude exhibits substantially fewer Trigger-dependent success cases than Trigger-absent-only success cases, with 17 and 50 bugs, respectively.
This result suggests that Claude is less effective at exploiting explicit trigger testcase information to guide repair generation.

\subsubsection{Line Coverage Prediction}

\begin{table}[t]
\centering
\scriptsize
\caption{Distribution of code-structure-based understanding hallucinations among low-scoring line coverage predictions. Counts in the \# Predictions column are reported in the order of Claude / DeepSeek / GPT-5. The categories are not mutually exclusive.}
\label{tab:task2-execution-understanding-hallucination}
\renewcommand{\arraystretch}{1.3}
\setlength{\tabcolsep}{4pt} 
\begin{tabular}{l|l|c}
\hline
\textbf{Category}
& \textbf{Code Structure Example}
& \textbf{\# Predictions} \\
\hline

Branch
& \sccode{if}, \sccode{else}, \sccode{switch-case}, ...
& 25 / 11 / 3 \\

Exception Flow
& \sccode{try}, \sccode{catch}, ...
& 3 / 2 / 1 \\

Return Statement
& \sccode{return}
& 2 / 0 / 0 \\

Assignment Statement
& Variable assignments and state updates
& 0 / 0 / 2 \\
\hline

\textbf{Total}
& --
& 28 / 11 / 5 \\
\hline

\end{tabular}
\end{table}

RQ1 shows that plausible repairs are generally associated with higher line coverage prediction accuracy.
Here, we move beyond aggregate prediction scores and examine the program structures associated with low-scoring predictions.
We independently select buggy- and model-patched-program predictions with an F1 score of at most $0.5$ for manual analysis.
We use this operational threshold to focus on predictions that substantially diverge from the observed coverage, rather than those containing only minor line-level discrepancies.
In total, we annotate 29 buggy-program predictions and 15
model-patched-program predictions.
We treat the two program versions independently: if both predictions for the same bug satisfy the threshold, they are counted as two separate samples.
Table~\ref{tab:task2-execution-understanding-hallucination} summarizes the code structures associated with these low-scoring predictions.

For each selected prediction, we inspect the regions where the
model-predicted coverage diverges from the observed coverage and annotate the associated code structures.

$\bullet$~\textbf{Branch} refers to hallucinations involving conditional
control flow, where the model predicts the wrong branch or misses the branch
actually executed by the triggering testcases.
Typical structures include \code{if}, \code{else}, and \code{switch-case}.

$\bullet$~\textbf{Exception Flow} refers to hallucinations involving
exception-related control flow, where the model incorrectly predicts whether
an exception-handling path is executed.
Typical structures include \code{try} and \code{catch}.

$\bullet$~\textbf{Return Statement} refers to hallucinations involving
whether a return statement is executed, causing the model to incorrectly
predict where control exits a method.

$\bullet$~\textbf{Assignment Statement} refers to hallucinations involving
whether an assignment or state-update statement is executed.

Since one incorrect prediction may involve multiple code structures, these
categories are not mutually exclusive.
Branch-related hallucinations dominate across all three models, appearing in
25 of 28 Claude predictions, all 11 DeepSeek predictions, and 3 of 5 GPT-5
predictions.
Exception-flow hallucinations are considerably less frequent, while return
and assignment statements occur only in isolated predictions.
These results suggest that the primary difficulty in line coverage prediction
lies in identifying the correct conditional execution path under the
triggering testcases.

\subsubsection{Additional Testcase Generation}

\begin{table*}[t]
\centering
\tiny
\caption{Distribution of understanding hallucinations in the additional testcase generation task. Counts in the \# Cases column are reported in the order of Claude / DeepSeek / GPT-5.}
\label{tab:task3-testcase-understanding-hallucination}
\renewcommand{\arraystretch}{1.5}
\setlength{\tabcolsep}{4pt}
\begin{tabular}{
>{\raggedright\arraybackslash}p{2.1cm}|
l|
c
}
\hline
\textbf{Category}
& \textbf{Execution-based Condition}
& \textbf{\# Cases} \\
\hline

Non-existent Symbol Reference
& $T_{\text{buggy}}=\Uncompilable$
& 4 / 4 / 5 \\

Missing Import or Dependency
& $T_{\text{buggy}}=\Uncompilable$
& 3 / 12 / 7 \\

Duplicated Declaration
& $T_{\text{buggy}}=\Uncompilable$
& 5 / 2 / 3 \\

Missing Test Method Wrapper
& $T_{\text{buggy}}=\Uncompilable$
& 0 / 0 / 7 \\

Malformed Escape Sequences
& $T_{\text{buggy}}=\Uncompilable$
& 0 / 4 / 1 \\

Test-framework/ Language Version Mismatch
& $T_{\text{buggy}}=\Uncompilable$
& 5 / 1 / 1 \\

\hline

Incorrect Output Expectation
& $T_{\text{gt}}=\Notpass$ with incorrect expected output
& 11 / 11 / 9 \\

Incorrect Exception Expectation
& $T_{\text{gt}}=\Notpass$ with incorrect expected exception
& 2 / 1 / 0 \\

Compilable Unrelated Runtime Failure
& $T_{\text{buggy}}=\Notpass \wedge T_{\text{gt}}=\Notpass \wedge T_{\text{model}}\in\{\Notpass,\Uncompilable\}$
& 4 / 4 / 0 \\

Overfitting Testcase
& $T_{\text{buggy}}=\Notpass \wedge T_{\text{gt}}=\Notpass \wedge T_{\text{model}}=\Pass$
& 1 / 4 / 4 \\

\hline

Faulty Code Location Not Reached
& $T_{\text{buggy}}=\Pass \wedge T_{\text{gt}}=\Pass$
& 4 / 3 / 0 \\

Incomplete Bug-Triggering Condition
& $T_{\text{buggy}}=\Pass \wedge T_{\text{gt}}=\Pass$
& 27 / 9 / 10 \\

Under-Specified Assertion
& $T_{\text{buggy}}=\Pass \wedge T_{\text{gt}}=\Pass$
& 3 / 0 / 0 \\

\hline

Partial Test
& $T_{\text{buggy}}=\Notpass \wedge T_{\text{gt}}=\Pass \wedge T_{\text{model}}=\Pass \wedge T_{\text{patch}}=\Notpass$
& 3 / 6 / 4 \\

\hline

Others
& \All
& 0 / 1 / 0 \\

\hline
\textbf{Total}
& --
& \textbf{72 / 62 / 51} \\
\hline
\end{tabular}
\end{table*}

We next examine how understanding hallucinations manifest in generated additional testcases.
Compared with triggering testcase identification and line coverage prediction, additional testcase generation introduces an extra generation step: the model must construct runnable test code in the project-specific testing style and encode an oracle for the expected post-repair behavior.
Thus, hallucinations in this task may arise not only from misunderstanding the bug-triggering behavior, but also from errors in testcase construction, oracle specification, or project-context grounding.

To further characterize these failures, we execute each generated testcase on three program versions, namely the original buggy program, the model-generated patched program, and the developer-written fixed program.
Let $T_{\text{buggy}}$, $T_{\text{model}}$, and $T_{\text{gt}}$ denote the execution outcome of the generated testcase on these three versions, respectively.
We further use $T_{\text{patch}}$ to denote the outcome of the model-generated patch on the developer-written test suite.
A generated testcase is considered valid if it fails on the original buggy program and passes on the developer-written fixed program, i.e., $T_{\text{buggy}}=\Fail \wedge T_{\text{gt}}=\Pass$.
Based on these execution outcomes and manual inspection, we classify the observed understanding hallucinations into fourteen categories and one Other group, as shown in Table~\ref{tab:task3-testcase-understanding-hallucination}. 
The case-specific annotations corresponding to this table will be released in our online replication package.

The first six categories capture hallucinations that typically make the generated testcases uncompilable. 

$\bullet$~\textbf{Non-existent Symbol Reference} and ~\textbf{Missing Import or Dependency} follow the same definitions as their corresponding repair hallucination categories introduced earlier.
The difference is that these hallucinations occur in the generated testcase rather than in the generated patch in this task.

$\bullet$~\textbf{Duplicated Declaration} refers to an understanding hallucination in which the generated testcases introduce declarations that conflict with existing test code.
This may include duplicated test method names or duplicated helper methods.

$\bullet$~\textbf{Missing Test Method Wrapper} refers to an understanding hallucination in which the generated testcase contains one or more test statements but does not enclose them within a valid test method declaration, causing compilation errors.

\begin{figure}[t]
\centering
\includegraphics[width=0.9\linewidth]{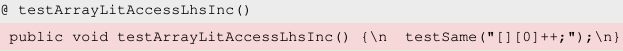}
\caption{Example of malformed escape sequences in the testcase generated for Closure-161.}
\label{fig:malformed-escape-closure161}
\end{figure}

$\bullet$~\textbf{Malformed Escape Sequences} refers to an understanding hallucination in which escape sequences in the generated testcase are incorrectly produced or preserved during output parsing, such as escaped quotation marks or newline sequences being written literally into the source file, thereby violating the target language syntax and causing compilation errors.
For example, in the testcase generated for Closure-161 by Claude, the complete test method is written on one physical line with the intended line breaks retained as literal \code{\textbackslash n} sequences, as shown in Figure~\ref{fig:malformed-escape-closure161}.
Similarly, generated testcases for Gson-11 and Jsoup-27 by GPT-5 contain unnecessary backslashes before the opening and closing quotation marks of Java string literals, resulting in illegal-character and unclosed-string compilation errors.

$\bullet$~\textbf{Test-framework/Language Version Mismatch} refers to an understanding hallucination in which the generated testcases use testing APIs, annotations, or language features that are incompatible with the target project.
For example, in the testcase generated by Claude for JacksonDatabind-97, the model introduces an annotation-style \code{@Test} method, but the target test context does not recognize the \code{Test} annotation, causing compilation failure.

The next four categories capture hallucinations that typically produce testcases failing on the developer-written fixed program.
Such outcomes suggest that the generated testcases are not aligned with the developer-intended repair semantics, indicating that the model does not correctly understand what behavior the developer fix is supposed to preserve or produce.

$\bullet$~\textbf{Incorrect Output Expectation} refers to understanding hallucinations in which the generated testcases assert an expected output, return value, or object state that is inconsistent with the developer-written fixed behavior.

$\bullet$~\textbf{Incorrect Exception Expectation} refers to understanding hallucinations in which the generated testcases may encode an incorrect expectation about exception behavior, such as cases where the testcase incorrectly predicts whether an exception should be thrown, or predicts the wrong exception type.

$\bullet$~\textbf{Compilable Unrelated Runtime Failure} refers to understanding hallucinations in which the generated testcases compile successfully, but fail at runtime due to errors unrelated to the target bug, such as invalid test setup, missing runtime context, or malformed inputs.
As a result, these testcases fail because the generated test itself is incorrectly constructed.

$\bullet$~\textbf{Overfitting Testcase} refers to understanding hallucinations in which the generated testcase fails on the original buggy program and passes on the model-generated patched program, but fails on the developer-written fixed program.
This indicates that the testcase is consistent with the model's own repair, but inconsistent with the developer-intended repair semantics.

The following three categories capture hallucinations that typically produce executable testcases passing on both the original buggy program and the developer-written fixed program, meaning they lack the ability to distinguish the buggy behavior from the intended repaired behavior.

$\bullet$~\textbf{Faulty Code Location Not Reached} refers to understanding hallucinations in which the generated testcases execute a path that bypasses the faulty code locations.
For example, the generated testcases may take a different branch from the bug-triggering branch and instead exercise a branch that is already correctly handled by the buggy program.

\begin{figure}[t]
\centering

\begin{subfigure}[t]{0.48\textwidth}
\centering
\includegraphics[width=\linewidth]{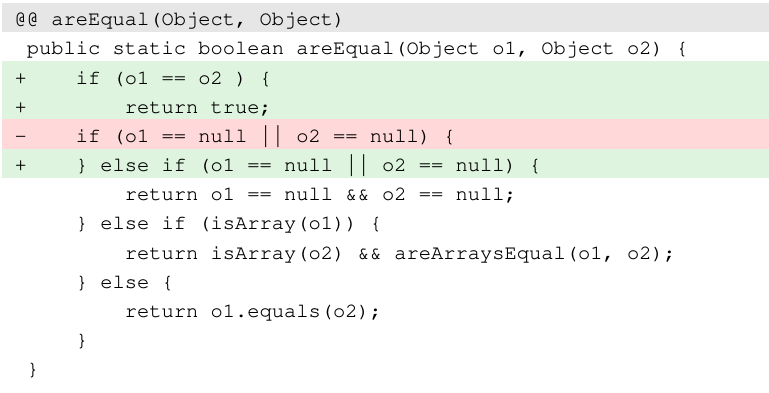}
\caption{Developer diff}
\label{fig:mockito22-developer-diff}
\end{subfigure}
\hfill
\begin{subfigure}[t]{0.48\textwidth}
\centering
\includegraphics[width=\linewidth]{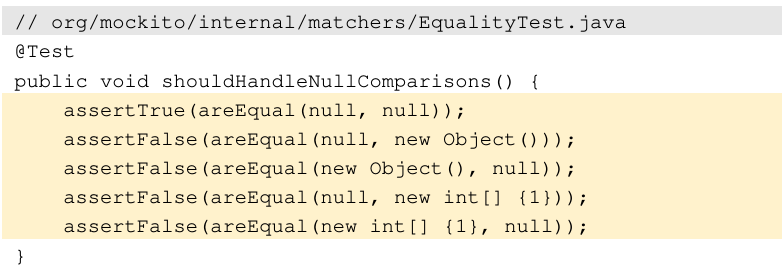}
\caption{Generated testcase}
\label{fig:mockito22-generated-testcase}
\end{subfigure}

\caption{Example of incomplete bug-triggering condition in the testcase generated by Claude for Mockito-22 in the additional testcase generation task.}
\label{fig:incomplete-trigger-mockito22}
\end{figure}

$\bullet$~\textbf{Incomplete Bug-Triggering Condition} refers to hallucinations in which the generated testcase may reach the faulty code location but omits the concrete input conditions needed to expose the bug, such as boundary values, configuration options, or input combinations.
For example, Figure~\ref{fig:incomplete-trigger-mockito22} shows a testcase generated by Claude for Mockito-22.
The developer fix adds an early reference-equality check, \sccode{o1 == o2}, in \sccode{areEqual}.
To expose this bug, a testcase needs to compare two references to the same non-null object.
However, the generated testcase only checks null-comparison cases, which are already handled correctly by the buggy program.
Thus, although the testcase reaches the relevant method, it omits the condition required to expose the bug.

$\bullet$~\textbf{Under-Specified Assertion} refers to an understanding hallucination in which the generated testcases may encode assertions that are too weak to distinguish the buggy behavior from the intended repaired behavior.
For example, the intended repair requires the testcase to distinguish two different outputs or observable behaviors, but the generated assertion is satisfied by both.
As a result, the testcase may pass on both the original buggy program and the developer-written fixed program.

The final category captures a special case among generated testcases that satisfy the usual validity condition.
In most cases, if a generated testcase fails on the original buggy program and passes on the developer-written fixed program, i.e., $T_{\text{buggy}}=\Fail \wedge T_{\text{gt}}=\Pass$, we treat it as a valid testcase.
However, this criterion may still overestimate testcase quality when the generated testcase only captures a partial aspect of the intended repair semantics.

$\bullet$~\textbf{Partial Test} refers to an understanding hallucination in which the generated testcase only tests part of the relevant code affected by the bug or the developer-written fix.

$\bullet$~\textbf{Others} includes rare understanding hallucinations that do not fit the preceding categories.
For example, in the testcase generated by DeepSeek for JacksonDatabind-27, a static helper class is declared inside a test method, which is illegal in Java and causes compilation failure.

Table~\ref{tab:task3-testcase-understanding-hallucination} reports 185 observed understanding hallucinations across the three models.
The most prevalent category is incomplete bug-triggering condition, accounting for 46 cases (24.9\%).
This suggests that models often capture the general bug context but omit the concrete boundary values, configurations, or input combinations required to expose the faulty behavior.

Testcase construction errors are also common.
The six compilation-related categories account for 64 cases (34.6\%), indicating that generating project-compatible test code remains challenging.
In addition to correctly using project-specific symbols, dependencies, testing frameworks, and language features, models must generate complete test-method structures and correctly encoded source code.
Failures such as missing test method wrappers and malformed escape sequences show that otherwise meaningful test statements may still be unusable because their syntactic structure or textual representation is invalid.
Oracle-related categories, such as incorrect output expectation and incorrect exception expectation, further show that models may fail to encode the developer-intended fixed behavior even when the generated testcase is executable.

Notably, 56 generated testcases pass on both the original buggy program and the developer-written fixed program.
Although these testcases are invalid under our primary criterion because they do not expose the target bug, they may still provide some value as regression testcases.
Since the exercised behavior is accepted by both the buggy and developer-fixed versions, it should generally remain unaffected by the repair.
These non-triggering testcases may thus complement bug-revealing testcases by checking whether a patch inadvertently breaks previously correct behavior.

These categories share some similarities with the repair hallucination taxonomy in Table~\ref{tab:repair-hallucination-classification}.
In particular, both generated patches and generated testcases exhibit hallucinations related to non-existent symbols and missing imports or dependencies, suggesting that insufficient project-context grounding is a common source of hallucination across different APR artifacts.

However, additional testcase generation also reveals task-specific failure modes that are less visible in patch generation.
A generated testcase must not only compile and execute, but also encode inputs that trigger the bug and an oracle that matches the developer-intended fixed behavior.
Thus, categories such as incomplete bug-triggering condition and incorrect output or exception expectation point to additional challenges in causal understanding and oracle specification.
These observations point to broader factors that may contribute to hallucinations across APR tasks.

\rqbox{\textbf{Finding 4.}
Understanding hallucinations remain prevalent across all three APR artifact-generation tasks.
In triggering testcase identification, exact identification rates remain low across models, ranging from 23.0\% to 40.9\%, and mixed misidentification affects 33.3\%--47.6\% of bugs.
In line coverage prediction, branch-related structures appear in 39 of 44 manually inspected low-scoring predictions (88.6\%).
In additional testcase generation, incomplete bug-triggering conditions are the largest category, accounting for 46 of 185 observed hallucinations (24.9\%).}

\subsection{Contributing Factors to Hallucinations}
\label{sec:result_RQ3}
The previous two research questions show that hallucinations appear not only in final repair patches, but also in intermediate APR artifacts, including triggering testcase identification, line coverage prediction, and additional testcase generation.
In this section, we further investigate what factors may explain the emergence of these hallucinations.

\begin{table*}[t]
\centering
\scriptsize
\caption{Potential factors contributing to hallucinations in LLM-based APR, synthesized from repair hallucinations and intermediate understanding hallucinations.}
\label{tab:rq3-possible-causes}
\renewcommand{\arraystretch}{1.25}
\setlength{\tabcolsep}{4pt}
\begin{tabular}{p{0.23\textwidth}|p{0.37\textwidth}|p{0.33\textwidth}}
\hline
\textbf{Contributing Factor}
& \textbf{Hallucination Types}
& \textbf{Interpretation} \\
\hline

Insufficient Project-Context Grounding
&
\begin{tabular}[t]{@{}l@{}}
Non-existent Symbol Reference \\
Missing Import or Dependency \\
Duplicated Declaration \\
Test-framework/ \\
Language Version Mismatch \\
Incorrect API Usage
\end{tabular}
& The model fails to ground its generated patch or testcase in the concrete project context, including available symbols, imports, dependencies, test framework conventions, and API semantics. \\

\hline

Imprecise Causal Localization
&
\begin{tabular}[t]{@{}l@{}}
Incorrect Causal Localization \\
Spurious Identification \\
Mixed Misidentification \\
Faulty Code Location Not Reached
\end{tabular}
& The model captures some bug-related signal, but cannot precisely distinguish causal testcases or causal code locations from merely related but non-causal context. \\

\hline

Unstable Execution-Path Reasoning
&
\begin{tabular}[t]{@{}l@{}}
Branch \\
Exception Flow
\end{tabular}
& The model has difficulty predicting which branch, guard, or exception-handling path is actually exercised by the triggering testcase. \\

\hline
\end{tabular}
\end{table*}

Table~\ref{tab:rq3-possible-causes} summarizes three possible causes of hallucinations observed across the studied APR tasks.




\subsubsection{Insufficient Project-Context Grounding}

As summarized in Tables~\ref{tab:repair-hallucination-classification} and~\ref{tab:task3-testcase-understanding-hallucination}, many hallucinations appear to stem from insufficient project-context grounding.
This factor is most evident in uncompilable patches and testcases.
Categories such as \emph{Non-existent Symbol Reference}, \emph{Missing Import or Dependency}, \emph{Duplicated Declaration}, and \emph{Test-framework/Language Version Mismatch} indicate that models may generate code that is plausible in a generic programming context but incompatible with the symbols, dependencies, or testing conventions of the target project.
\emph{Incorrect API Usage} reflects a related problem, in which a model modifies a relevant location but misinterprets project-specific API semantics, argument order, or return-value meaning.

During the annotation of repair hallucinations, we first examined the developer-written fix to understand the intended repair semantics.
We observed that many \emph{Non-existent Symbol Reference} cases occurred when the intended repair relied on a new helper method or variable introduced outside the buggy functions provided to the model.
Such repairs require the model to reconstruct not only the modification within the given function, but also its relationship with project-level entities absent from the input context.
For example, in Closure-1, the developer-written fix introduces a new project-level variable named \code{removeGlobals}.
The model-generated patch instead references \code{removeGlobal}, which is undefined in the project and therefore causes a compilation error.
Although the generated identifier is semantically plausible and closely resembles the entity required by the intended repair, it is not faithfully grounded in the concrete project state.

These observations suggest that insufficient project-context grounding may contribute to a subset of hallucinations, particularly when repair-critical entities are absent from the input.
Existing APR agents attempt to mitigate this problem through repository exploration, structure-aware context retrieval, fault localization, and iterative compilation- or test-based validation~\cite{autocoderover,patchpilot}.
More recent approaches further enrich the repair context using repository history or hierarchical code documentation~\cite{hafixagent,reporepair}.
However, indiscriminately expanding the context may increase token consumption while introducing irrelevant information.
In future work, we plan to investigate context-selection strategies that retrieve sufficient repair-critical information while maintaining relevance and token efficiency.

\subsubsection{Imprecise Causal Localization}

As summarized in Tables~\ref{tab:repair-hallucination-classification},
\ref{tab:task1-trigger-understanding-hallucination}, and
\ref{tab:task3-testcase-understanding-hallucination}, several hallucination categories are associated with imprecise causal localization.
In triggering testcase identification, \emph{Spurious Identification} and \emph{Mixed Misidentification} indicate that models may recognize testcases related to the buggy component but fail to distinguish those that actually expose the target bug.
At the patch level, \emph{Incorrect Causal Localization}, the most prevalent repair hallucination category, occurs when the model modifies code related to the observed failure without addressing its actual root cause.
In additional testcase generation, \emph{Faulty Code Location Not Reached} reflects a complementary failure mode, in which the generated testcase exercises valid project behavior but bypasses the faulty code path.
These observations suggest that imprecise causal localization may contribute to hallucinations across different APR artifacts.

\subsubsection{Unstable Execution-Path Reasoning}

As shown in Table~\ref{tab:task2-execution-understanding-hallucination}, unstable execution-path reasoning appears to be another potential contributing factor.
Branch-related structures account for 39 of the 44 manually inspected low-scoring predictions (88.6\%), substantially more than exception-flow, return-statement, and assignment-statement structures.
In the \emph{Branch} category, models predict an incorrect branch or omit the branch actually exercised by the triggering testcase.
The \emph{Exception Flow} category similarly captures errors in determining whether exception-handling paths are executed.
These patterns indicate that LLMs struggle to reason accurately about branching execution behavior, particularly in determining which path is exercised by a given testcase.

\rqbox{\textbf{Finding 5.} Hallucinations in LLM-based APR arise from insufficient project-context grounding, imprecise causal localization, and unstable execution-path reasoning. This indicates that models still struggle to ground generated artifacts in the target project, distinguish truly bug-causing evidence, and reason about the concrete paths.}

\

\section{Discussion}
\label{sec:discussion}
\subsection{Implications}

\textbf{Evaluation should incorporate intermediate artifacts and broader validation evidence. }
Our results show that the quality of intermediate artifacts, including identified trigger testcases, predicted line coverage, and generated additional testcases, is associated with final repair outcomes. 
Traditional APR evaluation largely determines patch correctness based on developer-written test cases, which may classify overfitting patches as plausible when the available test suite provides insufficient behavioral coverage. 
Future evaluations of LLM-based APR systems should therefore assess not only whether the final patch passes the existing tests, but also whether the intermediate artifacts faithfully represent the buggy behavior and intended repair.
Evidence derived from these artifacts could be aggregated into a confidence score that estimates patch reliability.
Moreover, they could support stage-aware diagnosis by revealing whether a failure originates from testcase understanding, execution-path reasoning, or repair synthesis.
Their correctness scores could additionally contribute to a patch-confidence estimate for identifying repairs that require further testing or manual inspection.

\textbf{APR systems should strengthen project grounding, causal localization, and execution-path reasoning.}
Our taxonomy indicates that hallucinations are frequently associated with insufficient project-context grounding, imprecise causal localization, and unstable execution-path reasoning.
Future systems should therefore integrate LLM generation with repository-aware retrieval, program analysis, and other sources of repair evidence.
Before generation, systems could retrieve repair-relevant symbols, dependencies, API usages, and testing conventions, together with non-code artifacts such as requirements, formal specifications, issue descriptions, and change history.
Dynamic coverage, program slicing, and execution traces could further help distinguish causal testcases and program locations from evidence that is merely related to the bug.
Because branch-dependent behavior is a frequent source of execution hallucinations, APR systems should also explicitly reason about the predicates, input values, program states, and configuration settings that determine the executed path.

\textbf{Intermediate artifacts should be treated as verifiable behavioral claims rather than inherently reliable explanations.}
Prior APR research has explored generated bug-reproduction tests for guiding and validating repair~\citep{cheng2025agenticbrt,cheng2026cogeneration}, iterative execution feedback for patch refinement~\citep{xia2024chatrepair,bouzenia2025repairagent}, runtime execution traces for fault diagnosis and patch validation~\citep{wu2026tracerepair}, and developer-facing explanations of generated repairs~\citep{lamahewage2025scholia}.
Our findings provide a complementary caution: model-generated intermediate artifacts may themselves contain hallucinations, including in cases where the corresponding patch passes all developer-written testcases.
Therefore, identified triggering testcases, predicted execution paths, and generated additional testcases should not be presented directly as evidence of patch correctness.
Instead, each artifact should be independently checked. 

\textbf{The hallucination taxonomy can guide targeted validation and model improvement.}
Different hallucination categories require different validation mechanisms.
Compilation and symbol-resolution checks can expose non-existent references and missing dependencies, whereas dynamic analysis, additional testing, and semantic comparison are needed for causal-localization errors, incorrect execution paths, and overfitting repairs.
The taxonomy could therefore support failure-specific validation pipelines rather than a single uniform patch checker.
It could also inform the construction of targeted training data.
Future work could investigate whether such taxonomy-guided training reduces specific forms of hallucination more effectively than simply increasing the amount of general repair data.

\textbf{Generated testcases can provide value beyond bug revelation.}
Although a bug-revealing testcase should fail on the buggy program and pass on the developer-written fixed program, a generated testcase that passes on both versions may still exercise behavior that should remain unchanged and thus serve as a regression-preserving testcase.
Conversely, a testcase that passes only on the model-patched program may expose patch-specific overfitting rather than validate the intended repaired behavior.
Future evaluations should therefore distinguish among bug-revealing, regression-preserving, patch-specific, and invalid testcases, assessing each according to the type of repair evidence it provides.

\subsection{Threats to Validity}
\label{sec:threats-validity}

\subsubsection{Internal validity.}
One threat to internal validity arises from the nondeterminism of LLM outputs.
Although we use the most deterministic decoding configuration supported by each provider API, not all APIs allow effective temperature control.
Repeated queries may therefore produce different outputs.
To ensure a consistent comparison, we collect one response per task instance using fixed prompt templates and keep the experimental settings consistent across tasks and models whenever supported.

Human interpretation may introduce internal bias, as annotators may apply category definitions inconsistently while the taxonomy evolves.
To mitigate this threat, two annotators independently inspect the sampled cases, iteratively refine the codebook through open coding, measure inter-annotator agreement, and resolve disagreements through discussion with two other experienced authors.

Potential implementation and parsing errors may affect the evaluation.
To mitigate this threat, we use a consistent pipeline, retain raw responses and execution logs and conduct all executions in isolated environments with clean project checkouts.

\subsubsection{Construct validity.}
Our study operationalizes repair hallucination and understanding hallucination using execution-based evidence and manual inspection.
For repair patches, we use developer-written tests to classify patches as \Pass, \Notpass, or \Uncompilable.
However, developer-written tests do not constitute a complete specification of the intended repair behavior.
A \Pass patch may therefore remain semantically incorrect or overfit the available tests.
To mitigate this limitation, we manually inspect sampled \Pass patches and compare them with the developer-written repairs.

The validity criterion for generated additional testcases is similarly limited.
We consider a testcase valid when it fails on the buggy version and passes on the developer-written fixed version.
Although this criterion establishes bug-revealing capability, it does not guarantee that the testcase captures the complete repair semantics, and testcases that pass on both versions may still provide regression-preserving value.
We therefore interpret this metric specifically as a measure of bug revelation rather than overall testcase usefulness and complement the automatic results with manual analysis of the observed failure patterns.

The contributing factors identified through manual analysis may also admit alternative explanations.
To reduce overinterpretation, we synthesize evidence across multiple hallucination categories and APR tasks and present these factors as empirically grounded potential contributors rather than established causal mechanisms.

\subsubsection{External validity.}
Our empirical evaluation is based on a single Java benchmark, \textsc{Defects4J}, and the observed results may differ for other datasets, programming languages, project ecosystems, or bug types.
To improve applicability beyond this setting, we design the evaluation framework around general APR artifacts and execution outcomes rather than Java-specific model behavior.
Given suitable test infrastructure and execution instrumentation, the same tasks can be instantiated for other programming languages and benchmarks.

We evaluate three LLMs, namely GPT-5, DeepSeek-R1, and Claude Sonnet 4.5, so the measured hallucination rates and category distributions may not generalize to other models or future versions.
To mitigate model-specific dependence, we select models from different providers and apply the same task definitions and evaluation procedures across them.
Moreover, the proposed tasks are model-agnostic and can be directly applied to newly released models by collecting the corresponding intermediate artifacts and evaluating them against execution-grounded oracles.

\subsubsection{Conclusion validity.}
The reported hallucination distributions are derived from samples rather than exhaustive annotation of all generated artifacts.
To reduce this threat, we use stratified sampling that covers every task--model combination and apply a common annotation procedure across all groups.

The observed relationship between intermediate-artifact quality and repair outcomes is correlational and does not establish that improving a particular artifact will necessarily improve repair success.
To avoid unsupported causal conclusions, we report the association consistently across tasks and models while explicitly distinguishing it from causation.
Our conclusions are therefore limited to showing that more accurate intermediate artifacts are generally associated with successful repairs, rather than claiming that they directly cause repair success.

\section{Conclusion}
\label{sec:conclusion}
This study investigates two forms of hallucination in LLM-based automated program repair: repair hallucination in final patches and understanding hallucination in intermediate APR artifacts.
We evaluate both through three tasks on 832 \textsc{Defects4J} bugs: triggering testcase identification, line coverage prediction, and additional testcase generation.

Hallucinations remain prevalent at both levels.
Only 21.0\%--55.9\% of generated patches pass the developer-written test suite, exact triggering-testcase identification rates range from 23.0\% to 40.9\%, and only 30.5\%--48.4\% of generated additional testcases are valid.
GPT-5 generally exhibits the lowest tendency toward hallucination, although accurate intermediate artifacts are not always accompanied by successful repairs.

Manual analysis of 812 repair outputs identifies repair hallucinations in 590 cases (72.7\%), with \emph{Incorrect Causal Localization} and \emph{Incorrect Repair Strategy} accounting for 45.9\% and 18.5\% of these hallucinations, respectively.
We further identify insufficient project-context grounding, imprecise causal localization, and unstable execution-path reasoning as potential contributing factors.
These findings motivate evaluating both final patches and intermediate artifacts and strengthening APR systems with better project grounding and execution evidence.

\section*{Statements and Declarations}
\subsection*{Funding}
This research is supported by the Ministry of Education, Singapore under its Academic Research Fund Tier 3 (Award ID: MOET32020-0004). Any opinions, findings and conclusions or recommendations expressed in this material are those of the author(s) and do not reflect the views of the Ministry of Education, Singapore.

\subsection*{Ethical approval}
Not applicable.

\subsection*{Informed consent}
Not applicable.

\subsection*{Author Contributions}
Xuemeng Cai conceived the study, designed the methodology, conducted the experiments, analyzed the results, performed the data annotation, and wrote the initial draft of the manuscript. 
Jiakun Liu contributed to the design of the methodology, resolved annotation conflicts, and validated the final taxonomy. 
Linhan Yang performed the data annotation and contributed to writing the manuscript. 
Wei Ma contributed to the conceptualization of the study and the design of the methodology. 
Lingxiao Jiang contributed to the conceptualization of the study and the design of the methodology, resolved annotation conflicts, validated the final taxonomy, and reviewed and revised the manuscript.

\subsection*{Data Availability Statement}

To support reproducibility, we provide a replication package containing the prompts, experimental results, annotation labels, and code.

The replication package is publicly available at:
\url{https://github.com/Cxm211/LLM_Hallucination}.

\subsection*{Conflict of Interest}
The authors have no competing interests to declare that are relevant to the content of this article.


\newpage
\balance
\bibliographystyle{spbasic}
\bibliography{ref}

@article{chen2021codex,
  title   = {Evaluating Large Language Models Trained on Code},
  author  = {Chen, Mark and Tworek, Jerry and Jun, Heewoo and Yuan, Qiming and
             de Oliveira Pinto, Henrique Ponde and Kaplan, Jared and Edwards, Harri and
             Burda, Yuri and Joseph, Nicholas and Brockman, Greg and Ray, Alex and
             Puri, Raul and Krueger, Gretchen and Petrov, Michael and Khlaaf, Heidy and
             Sastry, Girish and Mishkin, Pamela and Chan, Brooke and Gray, Scott and
             Ryder, Nick and Pavlov, Mikhail and Power, Alethea and Kaiser, Lukasz and
             Bavarian, Mohammad and Winter, Clemens and Tillet, Philippe and
             Such, Felipe Petroski and Cummings, Dave and Plappert, Matthias and
             Chantzis, Fotios and Barnes, Elizabeth and Herbert-Voss, Ariel and
             Guss, William Hebgen and Nichol, Alex and Paino, Alex and Tezak, Nikolas and
             Tang, Jie and Babuschkin, Igor and Balaji, Suchir and Jain, Shantanu and
             Saunders, William and Hesse, Christopher and Carr, Andrew N. and
             Leike, Jan and Achiam, Josh and Misra, Vedant and Morikawa, Evan and
             Radford, Alec and Knight, Matthew and Brundage, Miles and Murati, Mira and
             Mayer, Katie and Welinder, Peter and McGrew, Bob and Amodei, Dario and
             McCandlish, Sam and Sutskever, Ilya and Zaremba, Wojciech},
  journal = {CoRR},
  volume  = {abs/2107.03374},
  year    = {2021},
  url     = {https://arxiv.org/abs/2107.03374}
}

@inproceedings{
    jimenez2024swebench,
    title={{SWE}-bench: Can Language Models Resolve Real-world Github Issues?},
    author={Carlos E Jimenez and John Yang and Alexander Wettig and Shunyu Yao and Kexin Pei and Ofir Press and Karthik R Narasimhan},
    booktitle={The Twelfth International Conference on Learning Representations},
    year={2024},
    url={https://openreview.net/forum?id=VTF8yNQM66}
}

@inproceedings{lomshakov2024proconsul,
  title     = {{ProConSuL}: Project Context for Code Summarization with {LLMs}},
  author    = {Lomshakov, Vadim and Podivilov, Andrey and Savin, Sergey and
               Baryshnikov, Oleg and Lisevych, Alena and Nikolenko, Sergey I.},
  booktitle = {Proceedings of the 2024 Conference on Empirical Methods in Natural Language Processing: Industry Track},
  pages     = {866--880},
  year      = {2024},
  publisher = {Association for Computational Linguistics},
  url       = {https://aclanthology.org/2024.emnlp-industry.65/}
}

@inproceedings{xia2024chatrepair,
  title     = {Automated Program Repair via Conversation: Fixing 162 out of 337 Bugs for \$0.42 Each using {ChatGPT}},
  author    = {Xia, Chunqiu Steven and Zhang, Lingming},
  booktitle = {Proceedings of the 33rd ACM SIGSOFT International Symposium on Software Testing and Analysis},
  pages     = {819--831},
  year      = {2024},
  publisher = {Association for Computing Machinery},
  doi       = {10.1145/3650212.3680323}
}

@inproceedings{ribeiro2023llm4apr,
  author    = {Francisco Ribeiro and Jos{\'e} Nuno Macedo and Kanae Tsushima and Jo{\~a}o Saraiva},
  title     = {Large Language Models for Automated Program Repair},
  booktitle = {Companion Proceedings of the 2023 ACM SIGPLAN International Conference on Systems, Programming, Languages, and Applications: Software for Humanity},
  year      = {2023},
  doi       = {10.1145/3618305.3623587},
  url       = {https://dl.acm.org/doi/10.1145/3618305.3623587}
}

@inproceedings{yin2024thinkrepair,
  author    = {Xin Yin and Chao Ni and Shaohua Wang and Zhenhao Li and Limin Zeng and Xiaohu Yang},
  title     = {ThinkRepair: Self-Directed Automated Program Repair},
  booktitle = {Proceedings of the 33rd ACM SIGSOFT International Symposium on Software Testing and Analysis},
  year      = {2024},
  doi       = {10.1145/3650212.3680359},
  url       = {https://dl.acm.org/doi/10.1145/3650212.3680359}
}

@inproceedings{bouzenia2025repairagent,
  author    = {Islem Bouzenia and Premkumar Devanbu and Michael Pradel},
  title     = {RepairAgent: An Autonomous, LLM-Based Agent for Program Repair},
  booktitle = {Proceedings of the 47th IEEE/ACM International Conference on Software Engineering},
  year      = {2025},
  doi       = {10.1109/ICSE55347.2025.00157},
  url       = {https://dl.acm.org/doi/10.1109/ICSE55347.2025.00157}
}

@article{legoues2012genprog,
  author  = {Claire Le Goues and ThanhVu Nguyen and Stephanie Forrest and Westley Weimer},
  title   = {GenProg: A Generic Method for Automatic Software Repair},
  journal = {IEEE Transactions on Software Engineering},
  volume  = {38},
  number  = {1},
  pages   = {54--72},
  year    = {2012},
  doi     = {10.1109/TSE.2011.104}
}

@inproceedings{kim2013par,
  author    = {Dongsun Kim and Jaechang Nam and Jaewoo Song and Sunghun Kim},
  title     = {Automatic Patch Generation Learned from Human-Written Patches},
  booktitle = {Proceedings of the 2013 International Conference on Software Engineering},
  pages     = {802--811},
  year      = {2013},
  doi       = {10.1109/ICSE.2013.6606626}
}

@inproceedings{long2015spr,
  author    = {Fan Long and Martin Rinard},
  title     = {Staged Program Repair with Condition Synthesis},
  booktitle = {Proceedings of the 10th Joint Meeting on Foundations of Software Engineering},
  pages     = {166--178},
  year      = {2015},
  doi       = {10.1145/2786805.2786811}
}

@inproceedings{liu2019tbar,
  author    = {Kui Liu and Anil Koyuncu and Dongsun Kim and Tegawend{\'e} F. Bissyand{\'e}},
  title     = {TBar: Revisiting Template-Based Automated Program Repair},
  booktitle = {Proceedings of the 28th ACM SIGSOFT International Symposium on Software Testing and Analysis},
  year      = {2019},
  doi       = {10.1145/3293882.3330577},
  url       = {https://doi.org/10.1145/3293882.3330577}
}

@article{Huang2025,
author = {Huang, Lei and Yu, Weijiang and Ma, Weitao and Zhong, Weihong and Feng, Zhangyin and Wang, Haotian and Chen, Qianglong and Peng, Weihua and Feng, Xiaocheng and Qin, Bing and Liu, Ting},
title = {A Survey on Hallucination in Large Language Models: Principles, Taxonomy, Challenges, and Open Questions},
year = {2025},
issue_date = {March 2025},
publisher = {Association for Computing Machinery},
address = {New York, NY, USA},
volume = {43},
number = {2},
issn = {1046-8188},
url = {https://doi.org/10.1145/3703155},
doi = {10.1145/3703155},
journal = {ACM Trans. Inf. Syst.},
month = jan,
articleno = {42},
numpages = {55}
}

@article{zhang2025survey,
author = {Zhang, Ziyao and Wang, Chong and Wang, Yanlin and Shi, Ensheng and Ma, Yuchi and Zhong, Wanjun and Chen, Jiachi and Mao, Mingzhi and Zheng, Zibin},
title = {LLM Hallucinations in Practical Code Generation: Phenomena, Mechanism, and Mitigation},
year = {2025},
issue_date = {July 2025},
publisher = {Association for Computing Machinery},
address = {New York, NY, USA},
volume = {2},
number = {ISSTA},
url = {https://doi.org/10.1145/3728894},
doi = {10.1145/3728894},
journal = {Proc. ACM Softw. Eng.},
month = jun,
articleno = {ISSTA022},
numpages = {23}
}

@article{bang2023multitask,
  author  = {Bang, Yejin and Cahyawijaya, Samuel and Lee, Nayeon and Dai, Wenliang and Su, Dan and Wilie, Bryan and Lovenia, Holy and Ji, Ziwei and Yu, Tiezheng and Chung, Willy and Do, Quyet V. and Xu, Yan and Fung, Pascale},
  title   = {A Multitask, Multilingual, Multimodal Evaluation of {ChatGPT} on Reasoning, Hallucination, and Interactivity},
  journal = {CoRR},
  volume  = {abs/2302.04023},
  year    = {2023},
  url     = {https://arxiv.org/abs/2302.04023}
}

@article{guerreiro2023hallucinations,
  author  = {Guerreiro, Nuno Miguel and Alves, Duarte M. and Waldendorf, Jonas and Haddow, Barry and Birch, Alexandra and Colombo, Pierre and Martins, Andr{\'e} F. T.},
  title   = {Hallucinations in Large Multilingual Translation Models},
  journal = {CoRR},
  volume  = {abs/2303.16104},
  year    = {2023},
  url     = {https://arxiv.org/abs/2303.16104}
}

@inproceedings{yang2025coast,
  author    = {Yang, Weiqing and Wang, Hanbin and Liu, Zhenghao and Li, Xinze and Yan, Yukun and Wang, Shuo and Gu, Yu and Yu, Minghe and Liu, Zhiyuan and Yu, Ge},
  title     = {COAST: Enhancing the Code Debugging Ability of {LLMs} through Communicative Agent Based Data Synthesis},
  booktitle = {Findings of the Association for Computational Linguistics: NAACL 2025},
  pages     = {2570--2585},
  year      = {2025},
  publisher = {Association for Computational Linguistics}
}

@inproceedings{defects4j,
author = {Just, Ren\'{e} and Jalali, Darioush and Ernst, Michael D.},
title = {Defects4J: a database of existing faults to enable controlled testing studies for Java programs},
year = {2014},
isbn = {9781450326452},
publisher = {Association for Computing Machinery},
address = {New York, NY, USA},
url = {https://doi.org/10.1145/2610384.2628055},
doi = {10.1145/2610384.2628055},
booktitle = {Proceedings of the 2014 International Symposium on Software Testing and Analysis},
pages = {437–440},
numpages = {4},
location = {San Jose, CA, USA},
series = {ISSTA 2014}
}

@misc{gpt5,
  title        = {Introducing GPT-5},
  author       = {{OpenAI}},
  year         = {2025},
  month        = aug,
  howpublished = {\url{https://openai.com/index/introducing-gpt-5/}},
  note         = {Accessed: 2026-05-10}
}

@misc{claude45,
  title        = {Introducing Claude Sonnet 4.5},
  author       = {{Anthropic}},
  year         = {2025},
  month        = sep,
  howpublished = {\url{https://www.anthropic.com/news/claude-sonnet-4-5}},
  note         = {Accessed: 2026-05-10}
}

@article{motwani2020quality,
  author  = {Motwani, Manish and Soto, Mauricio and Brun, Yuriy and Just, Ren{\'e} and Le Goues, Claire},
  title   = {Quality of Automated Program Repair on Real-World Defects},
  journal = {IEEE Transactions on Software Engineering},
  volume  = {48},
  number  = {2},
  pages   = {637--661},
  year    = {2022},
  doi     = {10.1109/TSE.2020.2984918},
  url     = {https://doi.org/10.1109/TSE.2020.2984918}
}

@inproceedings{petke2024patchoverfitting,
  author    = {Petke, Justyna and Martinez, Matias and Kechagia, Maria and Aleti, Aldeida and Sarro, Federica},
  title     = {The Patch Overfitting Problem in Automated Program Repair: Practical Magnitude and a Baseline for Realistic Benchmarking},
  booktitle = {Companion Proceedings of the 32nd ACM International Conference on the Foundations of Software Engineering},
  year      = {2024},
  doi       = {10.1145/3663529.3663776},
  url       = {https://doi.org/10.1145/3663529.3663776}
}

@inproceedings{smith2015cure,
author = {Smith, Edward K. and Barr, Earl T. and Le Goues, Claire and Brun, Yuriy},
title = {Is the Cure Worse than the Disease? Overfitting in Automated Program Repair},
booktitle = {Proceedings of the 2015 10th Joint Meeting on Foundations of Software Engineering},
series = {ESEC/FSE 2015},
pages = {532--543},
year = {2015},
publisher = {Association for Computing Machinery},
doi = {10.1145/2786805.2786825}
}

@inproceedings{xin2017identifying,
author = {Xin, Qi and Reiss, Steven P.},
title = {Identifying Test-Suite-Overfitted Patches through Test Case Generation},
booktitle = {Proceedings of the 26th ACM SIGSOFT International Symposium on Software Testing and Analysis},
series = {ISSTA 2017},
pages = {226--236},
year = {2017},
publisher = {Association for Computing Machinery},
doi = {10.1145/3092703.3092718}
}

@inproceedings{long2016prophet,
  author    = {Long, Fan and Rinard, Martin},
  title     = {Automatic Patch Generation by Learning Correct Code},
  booktitle = {Proceedings of the 43rd Annual ACM SIGPLAN-SIGACT Symposium on Principles of Programming Languages},
  series    = {POPL '16},
  year      = {2016},
  pages     = {298--312},
  publisher = {ACM},
  doi       = {10.1145/2837614.2837617}
}

@article{martinez2017automatic,
  author  = {Martinez, Matias and Durieux, Thomas and Sommerard, Romain and Xuan, Jifeng and Monperrus, Martin},
  title   = {Automatic Repair of Real Bugs in Java: A Large-Scale Experiment on the Defects4J Dataset},
  journal = {Empirical Software Engineering},
  volume  = {22},
  number  = {4},
  pages   = {1936--1964},
  year    = {2017},
  doi     = {10.1007/s10664-016-9470-4}
}

@inproceedings{ye2022selfapr,
  author    = {Ye, He and Martinez, Matias and Luo, Xiapu and Zhang, Tao and Monperrus, Martin},
  title     = {SelfAPR: Self-Supervised Program Repair with Test Execution Diagnostics},
  booktitle = {Proceedings of the 37th IEEE/ACM International Conference on Automated Software Engineering},
  series    = {ASE '22},
  year      = {2022},
  publisher = {ACM},
  doi       = {10.1145/3551349.3556926}
}

@inproceedings{xia2022alpharepair,
  author    = {Xia, Chunqiu Steven and Zhang, Lingming},
  title     = {Less Training, More Repairing Please: Revisiting Automated Program Repair via Zero-Shot Learning},
  booktitle = {Proceedings of the 30th ACM Joint European Software Engineering Conference and Symposium on the Foundations of Software Engineering},
  series    = {ESEC/FSE '22},
  year      = {2022},
  publisher = {ACM},
  doi       = {10.1145/3540250.3549101}
}

@inproceedings{xia2023plmrepair,
  author    = {Xia, Chunqiu Steven and Wei, Yuxiang and Zhang, Lingming},
  title     = {Automated Program Repair in the Era of Large Pre-Trained Language Models},
  booktitle = {Proceedings of the 45th IEEE/ACM International Conference on Software Engineering},
  series    = {ICSE '23},
  year      = {2023},
  pages     = {1482--1494},
  publisher = {IEEE},
  doi       = {10.1109/ICSE48619.2023.00129}
}

@misc{eghbali2024dehallucinator,
  author       = {Eghbali, Aryaz and Pradel, Michael},
  title        = {De-Hallucinator: Mitigating LLM Hallucinations in Code Generation Tasks via Iterative Grounding},
  year         = {2024},
  eprint       = {2401.01701},
  archivePrefix = {arXiv},
  primaryClass = {cs.SE},
  url          = {https://arxiv.org/abs/2401.01701}
}

@misc{lee2025hallucination,
  author       = {Lee, Yunseo and Song, John Youngeun and Kim, Dongsun and Kim, Jindae and Kim, Mijung and Nam, Jaechang},
  title        = {Hallucination by Code Generation LLMs: Taxonomy, Benchmarks, Mitigation, and Challenges},
  year         = {2025},
  eprint       = {2504.20799},
  archivePrefix = {arXiv},
  primaryClass = {cs.SE},
  url          = {https://arxiv.org/abs/2504.20799}
}

@article{tambon2025bugs,
  author  = {Tambon, Florian and Moradi Dakhel, Arghavan and Nikanjam, Amin and Khomh, Foutse and Desmarais, Michel C. and Antoniol, Giuliano},
  title   = {Bugs in Large Language Models Generated Code: An Empirical Study},
  journal = {Empirical Software Engineering},
  year    = {2025},
  doi     = {10.1007/s10664-025-10614-4},
  url     = {https://doi.org/10.1007/s10664-025-10614-4}
}

@article{liu2024agents,
  author  = {Liu, Jiawei and Wang, Kaixin and Chen, Yuchuan and Peng, Xin and Chen, Zhenchang and Liu, Yang},
  title   = {Large Language Model-Based Agents for Software Engineering: A Survey},
  journal = {ACM Transactions on Software Engineering and Methodology},
  year    = {2024},
  doi     = {10.1145/3796507},
  url     = {https://doi.org/10.1145/3796507}
}

@article{xia2025agentless,
  author  = {Xia, Chunqiu Steven and Deng, Yinlin and Dunn, Soren and Zhang, Lingming},
  title   = {Demystifying LLM-Based Software Engineering Agents},
  journal = {Proceedings of the ACM on Software Engineering},
  year    = {2025},
  doi     = {10.1145/3715754},
  url     = {https://doi.org/10.1145/3715754}
}

@misc{bouzenia2025understandingagents,
  author        = {Bouzenia, Islem and Pradel, Michael},
  title         = {Understanding Software Engineering Agents: A Study of Thought-Action-Result Trajectories},
  year          = {2025},
  eprint        = {2506.18824},
  archivePrefix = {arXiv},
  primaryClass  = {cs.SE},
  url           = {https://arxiv.org/abs/2506.18824}
}

@article{deepseekr1,
  title         = {DeepSeek-R1: Incentivizing Reasoning Capability in LLMs via Reinforcement Learning},
  author        = {{DeepSeek-AI}},
  journal       = {arXiv preprint arXiv:2501.12948},
  year          = {2025},
  eprint        = {2501.12948},
  archivePrefix = {arXiv},
  primaryClass  = {cs.CL},
  url           = {https://arxiv.org/abs/2501.12948}
}

@misc{jacoco,
  title        = {{JaCoCo}: Java Code Coverage Library},
  author       = {{JaCoCo Team}},
  year         = {2025},
  howpublished = {\url{https://github.com/jacoco/jacoco}},
  note         = {Accessed: 2025-11-01}
}

@inproceedings{nguyen2013semfix,
  author    = {Hoang Duong Thien Nguyen and
               Dawei Qi and
               Abhik Roychoudhury and
               Satish Chandra},
  title     = {SemFix: Program Repair via Semantic Analysis},
  booktitle = {Proceedings of the 35th International Conference on Software
               Engineering},
  pages     = {772--781},
  publisher = {IEEE Computer Society},
  year      = {2013},
  doi       = {10.1109/ICSE.2013.6606623}
}

@misc{peng2025corrects,
      title={When "Correct" Is Not Safe: Can We Trust Functionally Correct Patches Generated by Code Agents?}, 
      author={Yibo Peng and James Song and Lei Li and Xinyu Yang and Mihai Christodorescu and Ravi Mangal and Corina Pasareanu and Haizhong Zheng and Beidi Chen},
      year={2025},
      eprint={2510.17862},
      archivePrefix={arXiv},
      primaryClass={cs.CR},
      url={https://arxiv.org/abs/2510.17862}, 
}

@misc{chen2025redteamingprogramrepair,
      title={Red Teaming Program Repair Agents: When Correct Patches can Hide Vulnerabilities}, 
      author={Simin Chen and Yixin He and Suman Jana and Baishakhi Ray},
      year={2025},
      eprint={2509.25894},
      archivePrefix={arXiv},
      primaryClass={cs.SE},
      url={https://arxiv.org/abs/2509.25894}, 
}

@article{gandhi2025whenagentsgoastray,
  title   = {When Agents Go Astray: Course-Correcting {SWE} Agents with {PRM}s},
  author  = {Gandhi, Shubham and
             Tsay, Jason and
             Ganhotra, Jatin and
             Kate, Kiran and
             Rizk, Yara},
  journal = {arXiv preprint arXiv:2509.02360},
  year    = {2025},
  doi     = {10.48550/arXiv.2509.02360}
}

@inproceedings{autocoderover,
  author    = {Yuntong Zhang and Haifeng Ruan and Zhiyu Fan and
               Abhik Roychoudhury},
  title     = {{AutoCodeRover}: Autonomous Program Improvement},
  booktitle = {Proceedings of the 33rd ACM SIGSOFT International Symposium
               on Software Testing and Analysis},
  series    = {ISSTA 2024},
  pages     = {1592--1604},
  year      = {2024},
  publisher = {Association for Computing Machinery},
  doi       = {10.1145/3650212.3680384},
  url       = {https://doi.org/10.1145/3650212.3680384}
}

@inproceedings{patchpilot,
  author    = {Hongwei Li and Yuheng Tang and Shiqi Wang and Wenbo Guo},
  title     = {{PatchPilot}: A Cost-Efficient Software Engineering Agent
               with Early Attempts on Formal Verification},
  booktitle = {Proceedings of the 42nd International Conference on
               Machine Learning},
  series    = {Proceedings of Machine Learning Research},
  volume    = {267},
  pages     = {35922--35941},
  year      = {2025},
  publisher = {PMLR},
  url       = {https://proceedings.mlr.press/v267/li25cf.html}
}

@article{hafixagent,
  author  = {Yu Shi and Hao Li and Bram Adams and Ahmed E. Hassan},
  title   = {{HAFixAgent}: History-Aware Automated Program Repair Agent},
  journal = {arXiv preprint arXiv:2511.01047},
  year    = {2025},
  eprint  = {2511.01047},
  archivePrefix = {arXiv},
  primaryClass  = {cs.SE},
  url     = {https://arxiv.org/abs/2511.01047}
}

@article{reporepair,
  author  = {Zhongqiang Pan and Chuanyi Li and Wenkang Zhong and Yi Feng
             and Bin Luo and Vincent Ng},
  title   = {{RepoRepair}: Leveraging Code Documentation for
             Repository-Level Automated Program Repair},
  journal = {arXiv preprint arXiv:2603.01048},
  year    = {2026},
  eprint  = {2603.01048},
  archivePrefix = {arXiv},
  primaryClass  = {cs.SE},
  url     = {https://arxiv.org/abs/2603.01048}
}

@article{cheng2025agenticbrt,
  author  = {Runxiang Cheng and Michele Tufano and J{\"u}rgen Cito and
             Jos{\'e} Cambronero and Pat Rondon and Renyao Wei and
             Aaron Sun and Satish Chandra},
  title   = {Agentic Bug Reproduction for Effective Automated Program Repair at Google},
  journal = {CoRR},
  volume  = {abs/2502.01821},
  year    = {2025},
  doi     = {10.48550/arXiv.2502.01821}
}

@inproceedings{cheng2026cogeneration,
  author    = {Runxiang Cheng and Michele Tufano and Jos{\'e} Cambronero and
               Renyao Wei and Sherry Shi and Grant Uy and Pat Rondon and
               Franjo Ivan{\v{c}}i{\'c}},
  title     = {Dynamic Cogeneration of Bug Reproduction Test in Agentic Program Repair},
  booktitle = {Proceedings of the 34th ACM International Conference on the
               Foundations of Software Engineering},
  year      = {2026}
}

@article{wu2026tracerepair,
  author  = {Jiaqing Wu and Tong Wu and Manqing Zhang and
             Yunwei Dong and Bo Shen},
  title   = {Runtime Execution Traces Guided Automated Program Repair
             with Multi-Agent Debate},
  journal = {CoRR},
  volume  = {abs/2604.02647},
  year    = {2026},
  doi     = {10.48550/arXiv.2604.02647}
}

@inproceedings{lamahewage2025scholia,
  author    = {Nethum Lamahewage and Nimantha Cooray and
               Ridwan Shariffdeen and Sandareka Wickramanayake and
               Nisansa de Silva},
  title     = {{SCHOLIA}: An {XAI} Framework for {APR}},
  booktitle = {Proceedings of the IEEE/ACM International Workshop on
               Automated Program Repair},
  pages     = {19--26},
  year      = {2025},
  doi       = {10.1109/APR66717.2025.00008}
}

@article{monperrus2018bibliography,
  author    = {Monperrus, Martin},
  title     = {Automatic Software Repair: A Bibliography},
  journal   = {ACM Computing Surveys},
  volume    = {51},
  number    = {1},
  articleno = {17},
  numpages  = {24},
  year      = {2018},
  publisher = {Association for Computing Machinery},
  doi       = {10.1145/3105906}
}

@article{legoues2019automated,
  author  = {Le Goues, Claire and Pradel, Michael and Roychoudhury, Abhik},
  title   = {Automated Program Repair},
  journal = {Communications of the ACM},
  volume  = {62},
  number  = {12},
  pages   = {56--65},
  year    = {2019},
  month   = dec,
  doi     = {10.1145/3318162}
}

@inproceedings{qi2015plausibility,
  author    = {Qi, Zichao and Long, Fan and Achour, Sara and Rinard, Martin C.},
  title     = {An Analysis of Patch Plausibility and Correctness for
               Generate-and-Validate Patch Generation Systems},
  booktitle = {Proceedings of the 24th International Symposium on
               Software Testing and Analysis},
  pages     = {24--36},
  year      = {2015},
  publisher = {Association for Computing Machinery},
  doi       = {10.1145/2771783.2771791}
}

@inproceedings{long2016searchspaces,
  author    = {Long, Fan and Rinard, Martin C.},
  title     = {An Analysis of the Search Spaces for Generate and Validate
               Patch Generation Systems},
  booktitle = {Proceedings of the 38th International Conference on
               Software Engineering},
  pages     = {702--713},
  year      = {2016},
  publisher = {Association for Computing Machinery},
  doi       = {10.1145/2884781.2884872}
}

@inproceedings{yang2017bettertests,
  author    = {Yang, Jinqiu and Zhikhartsev, Alexey and Liu, Yuefei and
               Tan, Lin},
  title     = {Better Test Cases for Better Automated Program Repair},
  booktitle = {Proceedings of the 11th Joint Meeting on Foundations of
               Software Engineering},
  pages     = {831--841},
  year      = {2017},
  publisher = {Association for Computing Machinery},
  doi       = {10.1145/3106237.3106274}
}

@inproceedings{lutellier2020coconut,
  author    = {Lutellier, Thibaud and Pham, Hung Viet and Pang, Lawrence and
               Li, Yitong and Wei, Moshi and Tan, Lin},
  title     = {{CoCoNuT}: Combining Context-Aware Neural Translation Models
               Using Ensemble for Program Repair},
  booktitle = {Proceedings of the 29th ACM SIGSOFT International Symposium
               on Software Testing and Analysis},
  year      = {2020},
  publisher = {Association for Computing Machinery},
  doi       = {10.1145/3395363.3397369}
}

@inproceedings{joshi2023ring,
  author    = {Joshi, Harshit and Cambronero Sanchez, Jos{\'e} and
               Gulwani, Sumit and Le, Vu and Radi{\v{c}}ek, Ivan and
               Verbruggen, Gust},
  title     = {Repair Is Nearly Generation: Multilingual Program Repair
               with {LLMs}},
  booktitle = {Proceedings of the AAAI Conference on Artificial Intelligence},
  volume    = {37},
  number    = {4},
  pages     = {5131--5140},
  year      = {2023},
  doi       = {10.1609/aaai.v37i4.25642}
}

@inproceedings{jin2023inferfix,
  author    = {Jin, Matthew and Shahriar, Syed and Tufano, Michele and
               Shi, Xin and Lu, Shuai and Sundaresan, Neel and
               Svyatkovskiy, Alexey},
  title     = {{InferFix}: End-to-End Program Repair with {LLMs}},
  booktitle = {Proceedings of the 31st ACM Joint European Software
               Engineering Conference and Symposium on the Foundations
               of Software Engineering},
  pages     = {1646--1656},
  year      = {2023},
  publisher = {Association for Computing Machinery},
  doi       = {10.1145/3611643.3613892}
}

@inproceedings{wang2023rapgen,
  author    = {Wang, Weishi and Wang, Yue and Joty, Shafiq and
               Hoi, Steven C. H.},
  title     = {{RAP-Gen}: Retrieval-Augmented Patch Generation with
               {CodeT5} for Automatic Program Repair},
  booktitle = {Proceedings of the 31st ACM Joint European Software
               Engineering Conference and Symposium on the Foundations
               of Software Engineering},
  pages     = {146--158},
  year      = {2023},
  publisher = {Association for Computing Machinery},
  doi       = {10.1145/3611643.3616256}
}

@article{silva2025repairllama,
  author  = {Silva, Andr{\'e} and Fang, Sen and Monperrus, Martin},
  title   = {{RepairLLaMA}: Efficient Representations and Fine-Tuned
             Adapters for Program Repair},
  journal = {IEEE Transactions on Software Engineering},
  volume  = {51},
  number  = {8},
  pages   = {2366--2380},
  year    = {2025},
  doi     = {10.1109/TSE.2025.3581062}
}

@article{wong2016faultlocalization,
  author  = {Wong, W. Eric and Gao, Ruizhi and Li, Yihao and
             Abreu, Rui and Wotawa, Franz},
  title   = {A Survey on Software Fault Localization},
  journal = {IEEE Transactions on Software Engineering},
  volume  = {42},
  number  = {8},
  pages   = {707--740},
  year    = {2016},
  doi     = {10.1109/TSE.2016.2521368}
}

@article{ji2023hallucination,
  author  = {Ji, Ziwei and Lee, Nayeon and Frieske, Rita and Yu, Tiezheng and
             Su, Dan and Xu, Yan and Ishii, Etsuko and Bang, Yejin and
             Madotto, Andrea and Fung, Pascale},
  title   = {Survey of Hallucination in Natural Language Generation},
  journal = {ACM Computing Surveys},
  volume  = {55},
  number  = {12},
  articleno = {248},
  numpages  = {38},
  year    = {2023},
  publisher = {Association for Computing Machinery},
  doi     = {10.1145/3571730}
}

@inproceedings{yang2024sweagent,
  author    = {Yang, John and Jimenez, Carlos E. and Wettig, Alexander and
               Lieret, Kilian and Yao, Shunyu and Narasimhan, Karthik and
               Press, Ofir},
  title     = {{SWE-agent}: Agent-Computer Interfaces Enable Automated
               Software Engineering},
  booktitle = {Advances in Neural Information Processing Systems},
  volume    = {37},
  year      = {2024},
  doi       = {10.52202/079017-1601}
}

\end{document}